\UseRawInputEncoding
\documentclass[aps,prx,twocolumn,superscriptaddress,showpacs]{revtex4-2}

\usepackage{graphicx}
\usepackage{dcolumn}
\usepackage{bm}
\usepackage{xcolor}
\usepackage{textgreek}

\newcommand{\ef}{$E_F$}
\newcommand{\kz}{$k_z$}
\newcommand{\bi}{Bi$_4$I$_4$}

\begin{document}

\title{Room-Temperature Topological Phase Transition in Quasi-One-Dimensional Material Bi$_4$I$_4$}

\author{Jianwei Huang}
\thanks{These authors contributed equally.}
\affiliation{Department of Physics and Astronomy, Rice University, Houston, TX 77005, USA}

\author{Sheng Li}
\thanks{These authors contributed equally.}
\affiliation{Department of Physics, The University of Texas at Dallas, 800 West Campbell Road, Richardson, Texas 75080-3021, USA}

\author{Chiho Yoon}
\thanks{These authors contributed equally.}
\affiliation{Department of Physics, The University of Texas at Dallas, 800 West Campbell Road, Richardson, Texas 75080-3021, USA}
\affiliation{Department of Physics and Astronomy, Seoul National University, Seoul 08826, Korea}

\author{Ji Seop Oh}
\affiliation{Department of Physics, University of California, Berkeley, Berkeley, California 94720, USA}
\affiliation{Department of Physics and Astronomy, Rice University, Houston, TX 77005, USA}

\author{Han Wu}
\affiliation{Department of Physics and Astronomy, Rice University, Houston, TX 77005, USA}

\author{Xiaoyuan Liu}
\affiliation{Department of Physics, The University of Texas at Dallas, 800 West Campbell Road, Richardson, Texas 75080-3021, USA}

\author{Nikhil Dhale}
\affiliation{Department of Physics, The University of Texas at Dallas, 800 West Campbell Road, Richardson, Texas 75080-3021, USA}

\author{Yan-Feng Zhou}
\affiliation{Department of Physics, The University of Texas at Dallas, 800 West Campbell Road, Richardson, Texas 75080-3021, USA}

\author{Yucheng Guo}
\affiliation{Department of Physics and Astronomy, Rice University, Houston, TX 77005, USA}

\author{Yichen Zhang}
\affiliation{Department of Physics and Astronomy, Rice University, Houston, TX 77005, USA}

\author{Makoto Hashimoto}
\affiliation{Stanford Synchrotron Radiation Lightsource, SLAC National Accelerator Laboratory, Menlo Park, California 94025, USA}

\author{Donghui Lu}
\affiliation{Stanford Synchrotron Radiation Lightsource, SLAC National Accelerator Laboratory, Menlo Park, California 94025, USA}

\author{Jonathan Denlinger}
\affiliation{Advanced Light Source, Lawrence Berkeley National Laboratory, Berkeley, California 94720, USA}

\author{Xiqu Wang}
\affiliation{Department of Chemistry, University of Houston, Houston, Texas 77204 USA}

\author{Chun Ning Lau}
\affiliation{Department of Physics, The Ohio State University, Columbus, OH 43210, USA}

\author{Robert J. Birgeneau}
\email{robertjb@berkeley.edu}
\affiliation{Department of Physics, University of California, Berkeley, Berkeley, California 94720, USA}
\affiliation{Materials Science Division, Lawrence Berkeley National Laboratory, Berkeley, California 94720,  USA}

\author{Fan Zhang}
\email{zhang@utdallas.edu}
\affiliation{Department of Physics, The University of Texas at Dallas, 800 West Campbell Road, Richardson, Texas 75080-3021, USA}

\author{Bing Lv}
\email{blv@utdallas.edu}
\affiliation{Department of Physics, The University of Texas at Dallas, 800 West Campbell Road, Richardson, Texas 75080-3021, USA}

\author{Ming Yi}
\email{mingyi@rice.edu}
\affiliation{Department of Physics and Astronomy, Rice University, Houston, TX 77005, USA}

\date{\today}

\begin{abstract}
Quasi-one-dimensional (1D) materials provide a superior platform for characterizing and tuning topological phases for two reasons: i) existence for multiple cleavable surfaces that enables better experimental identification of topological classification, and ii) stronger response to perturbations such as strain for tuning topological phases compared to higher dimensional crystal structures. In this paper, we present experimental evidence for a room-temperature topological phase transition in the quasi-1D material \bi, mediated via a first order structural transition between two distinct stacking orders of the weakly-coupled chains. Using high resolution angle-resolved photoemission spectroscopy on the two natural cleavable surfaces, we identify the high temperature $\beta$ phase to be the first weak topological insulator with gapless Dirac cones on the (100) surface and no Dirac crossing on the (001) surface, while in the low temperature $\alpha$ phase, the topological surface state on the (100) surface opens a gap, consistent with a recent theoretical prediction of a higher-order topological insulator beyond the scope of the established topological materials databases that hosts gapless hinge states. Our results not only identify a rare topological phase transition between first-order and second-order topological insulators but also establish a novel quasi-1D material platform for exploring unprecedented physics. 

\end{abstract}

\maketitle


\section{INTRODUCTION}

\begin{figure*}
\includegraphics[width=0.95\textwidth]{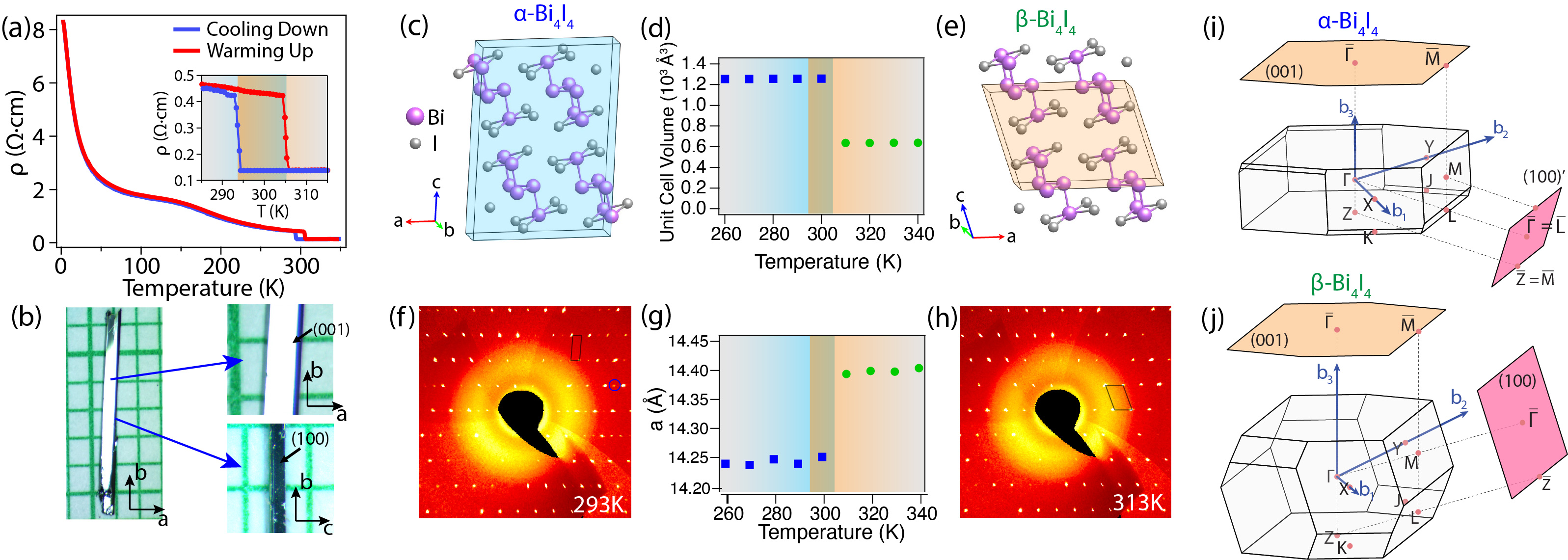}
\caption{\label{fig:Fig1} {\bf Crystal structure and sample characterization.} (a) Temperature dependent resistivity data for both cooling and warming curves. The inset is the enlarged view showing the hysteresis behavior of the first order transition. (b) Images of a \bi~crystal showing the cleavable (001) and (100) surfaces where the grid is 1 mm. (c) and (f) The real space crystal structure and the reciprocal space lattice from the h0l layer precession images integrated from X-ray diffraction data for the $\alpha$-\bi~phase. The shaded region is the structural unit cell. (d) and (g) The unit cell volume and refined lattice parameter $a$ from X-ray diffraction reflecting the $\alpha$ to $\beta$ structural transition.  (e,h) Same as (c,f) but for the $\beta$-\bi~phase.  (i) and (j) The bulk and projected surface Brillouin zones (BZ) of $\alpha$-\bi~and $\beta$-\bi.
}
\end{figure*}

Topological phases and associated phase transitions in quantum materials have garnered tremendous interest in recent years since the discovery of the two-dimensional (2D) quantum spin Hall (QSH) effect~\cite{Kane2005a, Konig2007}. Extending this concept to three dimensions (3D) led to the idea of topological insulator (TI), which hosts gapless boundary states. Within the class of TIs, there are two schemes by which the topological class can be further distinguished. The first scheme is the distinction between strong and weak TI~\cite{Fu2007, Moore2007}. While every surface of a strong TI possesses a gapless Dirac cone, such nontrivial surface states only appear on selected surfaces of a weak TI~\cite{Moore2007, Hasan2010, Qi2011, Liu2016}. The experimental identification between a strong and a weak TI therefore depends critically on the measurement of topological non-trivial surface states on multiple surfaces of a material, commonly probed via the technique of angle-resolved photoemission spectroscopy (ARPES). The challenge of such experimental efforts is that in natural materials, usually only a single preferred cleavage surface is experimentally accessible for ARPES. Most TIs that have been experimentally confirmed have been strong TIs, identified by a combination of a single-surface measurement and theoretical calculations, such as in the bismuth and antimony chalcogenides~\cite{Xia2009, Zhang2009, Chen2009, Hsieh2009}. Direct experimental verification of a weak TI, however, remains scarce due to the lack of materials with multiple cleavage surfaces where the existence of topologically nontrivial surface states could be probed~\cite{Liu2016, Yan2012, Rasche2013, Tang2014, Yang2014, Li2015}. The second scheme to classify further a TI is by its topological order, which is distinguished by the dimension of its gapless boundary states compared to the dimension of its bulk state. This realization led to the identification of a novel phase of matter dubbed the higher-order TI~\cite{Zhang2013, Po2017, Schindler2018, Yoon2020}, which are materials that host protected gapless states on boundaries of two or three spatial dimensions lower than its bulk. 

Interestingly, materials predicted to be candidate weak TIs are often in the vicinity of a topological critical point, where the exact topological classification depends sensitively upon the lattice parameters~\cite{Liu2016, Weng2014}. On the one hand, this produces an uncertainty in theoretical predictions of their topological properties, which makes experimental determination a necessity~\cite{Liu2016, Weng2014}. On the other hand, this renders an advantage that a topological phase transition between different topological phases can be easily induced by external perturbations such as strain and thermal effects~\cite{Weng2014, Liu2016, Mutch2019, Zhang2021}.

As a quasi-1D material, \bi~provides a superior platform not only to realize but also to be experimentally examined for the aforementioned distinct TI classes~\cite{Liu2016}. Consisting of stacked quasi-1D chains, \bi~naturally stabilizes in two crystal structures due to distinct stacking sequences, identified as $\alpha$-\bi~and $\beta$-\bi~\cite{VonSchnering1978}. The $\alpha$-phase is naturally stabilized at low temperatures and transitions into the $\beta$-phase via a first-order phase transition around 300 K. $\beta$-\bi~was theoretically predicted to be topologically nontrivial, in the vicinity of a transition point between weak and strong TIs~\cite{Liu2016, Autes2016}. Two subsequent ARPES studies reached conflicting conclusions~\cite{Autes2016, Noguchi2019}. While one study reports strong TI characteristics for a low temperature phase identified as $\beta$-\bi~\cite{Autes2016}, the other utilized a quenching method to stabilize the $\beta$-\bi~phase at low temperatures and identified it as a weak TI~\cite{Noguchi2019} while categorized the $\alpha$-\bi~phase as a trivial insulator with a distinct electronic structure~\cite{Noguchi2019}. Besides the controversy amongst the experimental reports, these results also differ from a recent theoretical work predicting the $\alpha$-phase to be a topologically nontrivial higher-order TI hosting helical hinge modes~\cite{Yoon2020}. While beyond the scope of the recently established topological materials database~\cite{Zhang2019a, Vergniory2019, Tang2019}, this predicted rare phase may be viewed as a QSH effect~\cite{Kane2005a, Konig2007} around the top or bottom surface of a 3D insulator, i.e., the time-reversal-invariant counterpart of the long-desired 3D quantum Hall effect~\cite{Halperin1987a, Tang2019a}. To resolve the aforementioned controversies, we systematically carried out electrical transport, X-ray diffraction, and ARPES measurements in combination with theoretical calculations to clarify the topological classification of $\alpha$-\bi~and $\beta$-\bi~as tuned by the natural structural transition via temperature.

\begin{figure*}
\includegraphics[width=0.9\textwidth]{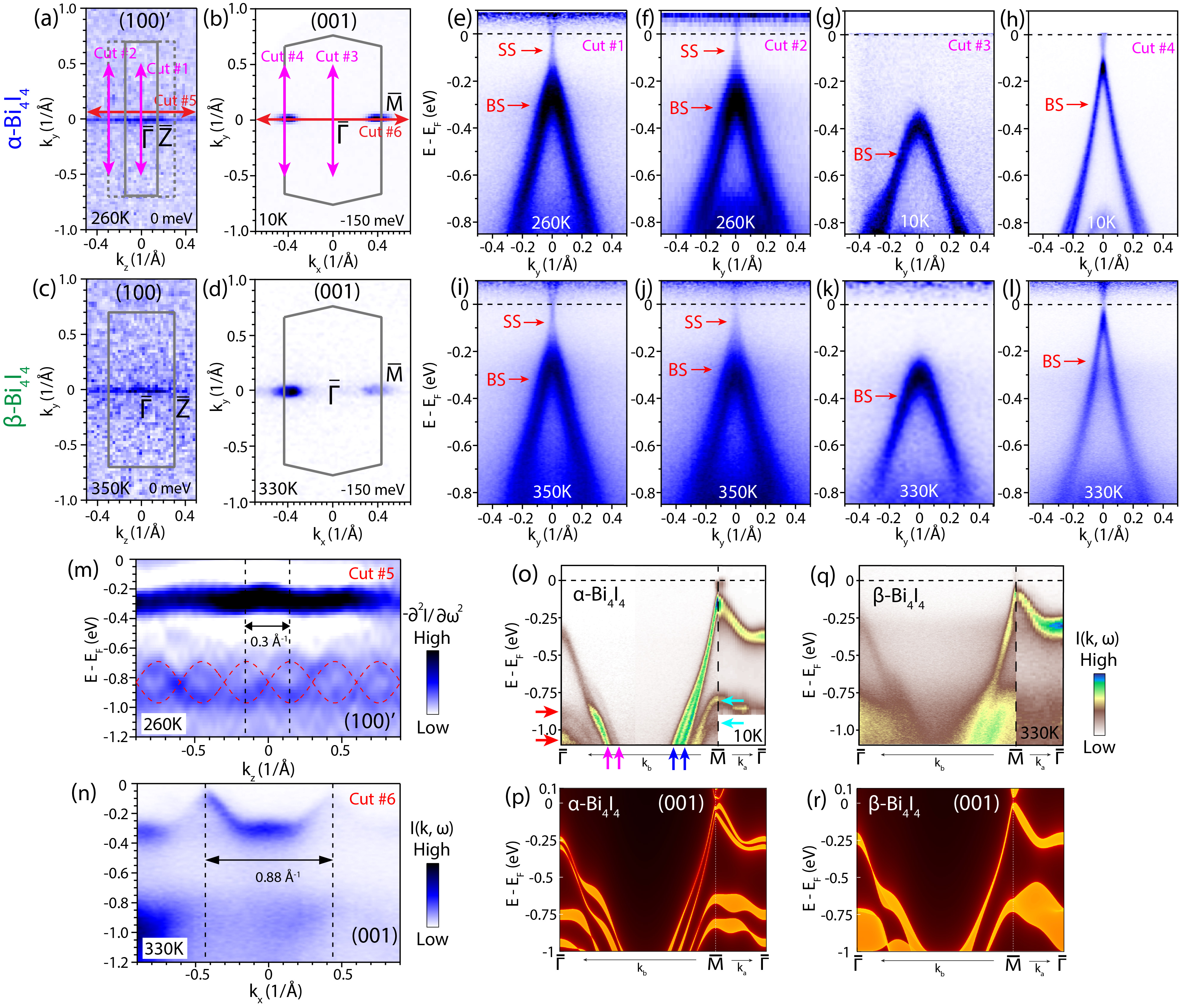}
\caption{\label{fig:Fig2} {\bf Electronic structure from different surfaces.} (a) Fermi surface mapping of $\alpha$-\bi~on the (100)' surface, with the surface BZ marked by gray rectangles (the solid one corresponds to $\alpha$-\bi~and the dashed one $\beta$-\bi). (b) Constant energy contour of $\alpha$-\bi~on the (001) surface. (e-h) Measured band dispersions of $\alpha$-\bi~along the cut \#1 to \#4 indicated in (a) and (b). Related surface states (SS) and bulk states (BS) are marked by arrows. (c-d) and (i-l) Same as the top row but for $\beta$-\bi. All band images are divided by the Fermi-Dirac function. (m) Band image along the cut \#5 except taken at $k_y=0$. Cut \#5 was drawn intentionally off-center to not block the Fermi surface image. The red dashed lines are sinusoidal fittings of the band image showing a periodicity of 0.3 \AA$^{-1}$ due to band doubling, which corresponds to the (100)' surface. (n) Band image along the cut \#6 indicated in (b). The black arrow shows a periodicity of 0.88 \AA$^{-1}$ which corresponds to the (001) surface. (o) Measured dispersions in $\alpha$-\bi~along high symmetry cuts. (p) Calculated bulk band structure of $\alpha$-\bi~projected onto the (001) surface (see Supplementary Materials). (q,r) Same as (o,p) but for $\beta$-\bi. A clear doubling of the bulk valence bands is evident from $\beta$-\bi~to $\alpha$-\bi~in both the measured and calculated dispersions. All measurement temperatures are as indicated.
}
\end{figure*}

\begin{figure*}
\includegraphics[width=0.9\textwidth]{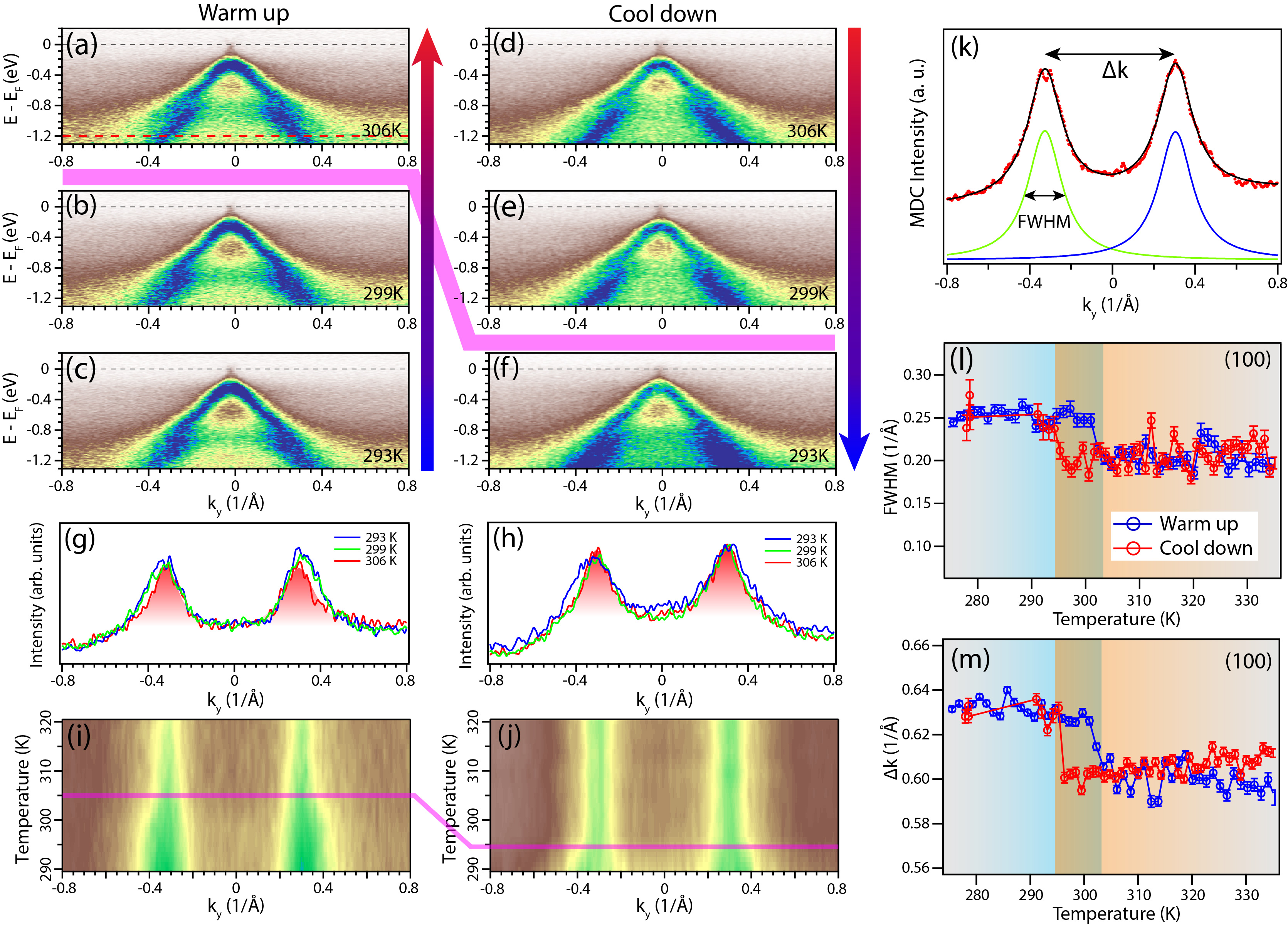}
\caption{\label{fig:Fig3} {\bf Bulk band doubling across the phase transition.} (a)-(c) Warm-up and (d)-(f) cool-down band images measured along cut \#1 indicated in Fig.~\ref{fig:Fig2}a at selected temperatures. (g)-(h) Corresponding MDCs at -1.2 eV during warm-up and cool-down, respectively. (i) Warm-up and (j) cool-down false-color images showing a more detailed temperature evolution of the MDCs indicated in (a). (k) Example of an MDC fitted by two Lorentzian peaks. (l) Full-width-at-half-max of the MDC fittings indicated in (k) as a function of temperature. The warm-up results are shifted 0.04 \AA~upwards. (m) The separation of the peak positions of the MDC fittings indicated in (k) as a function of temperature.
}
\end{figure*}

\section{Structural Transition between the $\alpha$- and $\beta$-phases}
Large \bi~single crystals were synthesized with the quasi-1D chains oriented along the $b$ axis, with large flat (001) and (100) surfaces (Fig.~\ref{fig:Fig1}) (see Supplementary Materials). The structural phase transition can be clearly identified from the temperature-dependent resistivity (Fig.~\ref{fig:Fig1}a) and heat capacity (Supplemental Fig. S1) measurements, where a hysteresis indicating a first order phase transition appears around 300 K. Below 270 K, the resistivity data show a typical semiconductor behavior with a broad hump at 150 K and are consistent with previous reports~\cite{Noguchi2019}. 

Structurally, the $\alpha$- and $\beta$-phases differ in the packing of the \bi~chains. The $\alpha$-\bi~has double-layered \bi~chains per unit cell (Fig.~\ref{fig:Fig1}c) whereas the $\beta$-\bi~has only a single-layer of \bi~chains per unit cell (Fig.~\ref{fig:Fig1}e). This subtle difference in stacking causes a large change of the $c$ lattice and the associated doubling of the unit cell volume (Fig.~\ref{fig:Fig1}d), as demonstrated by X-ray single crystal diffraction (Fig.~\ref{fig:Fig1}f, h). Contrasting the single crystal diffraction patterns for the reciprocal lattice $a^{\ast}$-$c^{\ast}$ planes taken in the $\alpha$-phase at 293 K (Fig.~\ref{fig:Fig1}f) and the $\beta$-phase at 313 K (Fig.~\ref{fig:Fig1}h), a near-doubling of the $c^{\ast}$ in reciprocal space from $\alpha$-\bi~to $\beta$-\bi~is visible, as seen by the additional peaks in Fig.~\ref{fig:Fig1}f marked by a circle. This structural change is well-captured in the temperature-dependent refined lattice parameters (Fig.~\ref{fig:Fig1}g and Fig. S2). The lattice parameters of the $\alpha$-phase are refined at 300 K to be $a = 14.251(2) \AA$, $b = 4.4304(5) \AA$, $c = 19.976(6) \AA$, and $\beta = 92.93(2)^{\circ}$, while those of the $\beta$-phase are refined at 310 K to be $a = 14.394(5) \AA$, $b = 4.4288(13) \AA$, $c = 10.494(5) \AA$, and $\beta = 107.96(5)^{\circ}$. 

The large facets on both the (001) ($>$ 0.5 mm) and (100) ($>$ 0.2 mm) surfaces are naturally cleavable, exposing flat surfaces that enable conventional ARPES studies on either surface using an incident beam spot size of 50 × 20 $\mu$m$^2$ (Fig.~\ref{fig:Fig1}b). We note the $\beta$-angle difference between $\alpha$-\bi~and $\beta$-\bi~crystal structures, where the natural cleavable side (100) surface of $\beta$-\bi~becomes the ($\bar{2}$01) surface in the $\alpha$-phase per definition in the true primitive unit cell (Fig.~\ref{fig:Fig1}c, e), we label it as (100)' in $\alpha$-\bi~(Fig.~\ref{fig:Fig1}i, j) to avoid confusion in the comparison with the (100) surface of the $\beta$-phase. 

Taking advantage of the rare multi-surface experimental access to \bi, we carried out ARPES measurements on both the (001) and (100) surfaces of the $\alpha$- and $\beta$-phases in their respective temperature regimes. We note that such two-surface ARPES measurement has also been desirable to identify unambiguously the topological characters of weak TIs and higher-order TIs. A comparison of the complete electronic structure of the two phases are summarized in Fig.~\ref{fig:Fig2}. Two main observations can be made: i) there is a strong distinction between the electronic structures measured on the (100) and (001) surfaces for both phases, and ii) the change between the two phases across the structural transition is subtle.

\section{Distinguishing the (001) and (100) Surfaces}
We first examine the distinction between the (100) and (001) surfaces, which is most clearly seen in the measured Fermi surfaces (FS). On the (100) surface, both phases exhibit a quasi-1D FS along \kz~(Fig.~\ref{fig:Fig2}a, c), revealing a rather weak interlayer coupling in the $c$-direction. The FS on the (001) surface, by contrast, consists of island-like bright spots for both phases (Fig.~\ref{fig:Fig2}b, d and Fig. S9), suggesting a stronger interlayer coupling along the $a$ axis compared to that of the $c$ axis. 
Furthermore, the measured sample surface orientation can be unambiguously distinguished by the periodicity of the observed band dispersions, which corresponds to the clearly distinct lattice parameters associated with the (100) and (001) surfaces. The $\alpha$-phase, for example, has a period of 0.3 \AA$^{-1}$ along the $k_z$ direction on the (100)' surface (Fig.~\ref{fig:Fig2}m) and 0.88 \AA$^{-1}$ along the $k_x$ direction on the (001) surface (Fig.~\ref{fig:Fig2}n). We note that we do not observe a drastic change of the fermiology across the phase transition, in contrast to the previous report that the quasi-1D FS only exists in the $\beta$-phase~\cite{Noguchi2019}. By clearly identifying single domains of each cleaved surface, we now clarify that the quasi-1D intensity and island-like intensity are due to the two different crystal surfaces measured rather than the two distinct structural phases.

We can furthermore examine the corresponding low-energy band dispersions. At the high symmetry points of both surfaces in both phases, an intense hole-like bulk band is observed below \ef, revealing a gap in the bulk state in all cases. The bulk nature of this band is confirmed by photon energy dependence as it disperses along \kz~(Fig. S6). However, two observations can be noted. Firstly, as the electronic structure is quasi-1D along $k_z$ on the (100) surfaces, the bands near the $\bar{{\Gamma}}$ and $\bar{Z}$ points are quite similar in both phases (Fig.~\ref{fig:Fig2}e,f,i,j) while those near the $\bar{{\Gamma}}$ and $\bar{M}$ points on the (001) surface are notably distinct. Secondly, while we observe a hole-like bulk valence band on all the cuts shown, within the bulk band gap on the (100) surface, a sharp Dirac surface state is observed for both the $\alpha$- and $\beta$-phases (Fig.~\ref{fig:Fig2}e,f,i,j). On the contrary, no Dirac surface state is observed inside the bulk band gap on the (001) surface of either phase (Fig.~\ref{fig:Fig2}g,h,k,l). At the $\bar{\Gamma}$-point, the bulk valence band tops at -0.3 eV. No surface state is observed. At the $\bar{M}$-point, the bulk valence band approaches \ef, with a visible gap near \ef~between the bulk valence and conduction bands.

\section{Electronic Reconstruction between the $\alpha$ and $\beta$ Phases}
Having distinguished the two surfaces, we now focus on the transition between the two structural phases. Overall, no dramatic change of the bulk band dispersions is observed across the phase transition. There is, however, a doubling in the number of bands from the $\beta$-phase to the $\alpha$-phase~\cite{Yoon2020} due to the doubling of the stacking layers in the unit cell. This band doubling is well-demonstrated by our theoretical calculations of the bulk bands projected onto the (001) surface for both the $\alpha$ and $\beta$ phases (Fig.~\ref{fig:Fig2}p, r and Supplementary Materials). While the band splitting is small due to the weak interlayer coupling, it is clearly resolved in our corresponding dispersions measured at low temperatures in the $\alpha$-phase (see pairs of arrows in Fig.~\ref{fig:Fig2}o,q as well as in a larger energy window shown in Fig. S3), consistent with the doubled stacking in the low temperature phase. 

Next, we demonstrate that this band doubling is indeed induced via the structural transition by systematically measuring across the structural transition near room temperature. We note that while it is challenging to resolved the small band splitting near the valence band top at elevated temperatures, this band doubling manifests as a broadening in the momentum distribution curves (MDCs) when a single band splits into two from the higher temperature $\beta$-phase to the lower temperature $\alpha$-phase. Clearly, this effect is in contrast to thermal broadening effects, which would have the opposite trend with respective to temperature. In Fig.~\ref{fig:Fig3}a-c, we plot the dispersions across the $\bar{{\Gamma}}$ point measured on (100) at selected temperatures. A clear abrupt broadening is indeed observed from the high temperature $\beta$-phase to the lower temperature $\alpha$-phase as exemplified by the MDCs at -1.2 eV (Fig.~\ref{fig:Fig3}i,j). Importantly, this abrupt change is accompanied by a hysteresis behavior revealed in the warm up (Fig.~\ref{fig:Fig3}i) and cool down (Fig.~\ref{fig:Fig3}j) processes. To quantify this behavior, we fit the MDCs and plot both their full-width-at-half-max (Fig.~\ref{fig:Fig3}l) and peak positions (Fig.~\ref{fig:Fig3}m), both of which show the hysteresis behavior reminiscent of that in the resistivity measurement, confirming the electronic structure’s response to the first order structural transition.

\begin{figure*}
\includegraphics[width=0.95\textwidth]{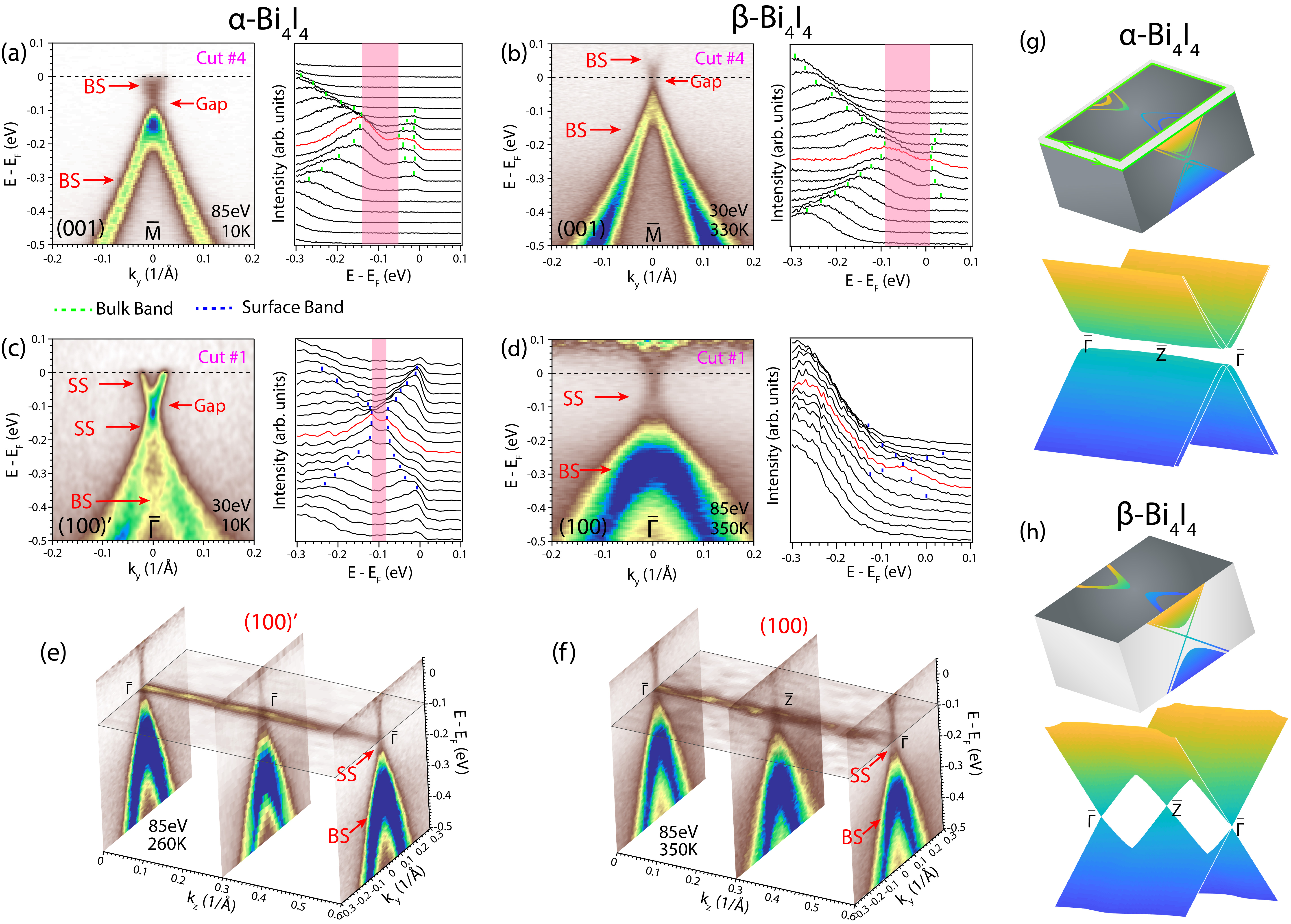}
\caption{\label{fig:Fig4} {\bf Surface states evolution across the phase transition.} (a) Band image along cut \#4 on the (001) surface indicated in Fig.~\ref{fig:Fig2}b of $\alpha$-\bi~as well as its corresponding EDC stacks. The red EDC corresponds to k = 0. The green dots track the peak positions of the bulk band which reveal a gap opening of $\sim$85 meV. (b) Same as (a) but of $\beta$-\bi. A gap opens at k = 0 of $\sim$100 meV. (c) Band image along cut \#1 on the (100)' surface indicated in Fig.~\ref{fig:Fig2}a of $\alpha$-\bi~as well as its corresponding EDC stacks. The blue dots track the surface state which reveal a $\sim$35 meV gap opening. (d) Same as (c) but of $\beta$-\bi. A gapless Dirac surface state is observed. (e) 3D view of the electronic structure on the (100)' surface of $\alpha$-\bi. A quasi-1D Dirac-like surface state is observed. (f) Same as (e) but of $\beta$-\bi. The quasi-1D Dirac surface state reveals the weak TI property. (g) Schematics (see Supplementary Materials) showing the higher-order TI state of $\alpha$-\bi~with a particular surface termination. Upper panel: all the bulk states and surface states are gapped, whereas a helical hinge state around the top surface exists in the surface state gap. Lower panel: the zoom-in (100)' surface state features a band doubling and a small gap. (h) Schematics (see Supplementary Materials) showing the weak TI state of $\beta$-\bi. Upper panel: the bulk states and (001) surface state are gapped, whereas there is a quasi-1D gapless Dirac surface state on the (100) surface. Lower panel: the zoom-in (100) surface state features two gapless Dirac cones. 
}
\end{figure*}

\section{Topological Characterization}
Having confirmed the phase transition as observed from the electronic structure, we examine the topological properties of the two phases. In Fig.~\ref{fig:Fig4} we present high resolution measurements near \ef~for both phases. We first examine the $\beta$-phase. On the (001) surface (Fig.~\ref{fig:Fig2}k, l and Fig.~\ref{fig:Fig4}b), we observe a clear gap in all the bulk gaps and no additional surface Dirac crossings. This can be confirmed from the energy distribution curve (EDC) stacks of the bands across $\bar{M}$ (Fig.~\ref{fig:Fig4}b), where the green dots track the bulk band dispersion and reveal a gap of $\sim$100 meV. Similar suppressed intensity can also be clearly identified in the MDC stacks (Fig. S7). On the (100) surface, a quasi-1D Dirac surface state exists inside the bulk band gap (Fig.~\ref{fig:Fig2}i, j and Fig.~\ref{fig:Fig4}d, f). Particularly at the $\bar{\Gamma}$ and $\bar{Z}$ points, the surface states are gapless within the resolution of measurements at 350 K. These combined measurements on both the (001) and (100) surfaces demonstrate that the high temperature $\beta$-\bi~is a weak TI where only selected surfaces possess an even number (two here) of gapless Dirac cones~\cite{Fu2007,Liu2016} (Fig.~\ref{fig:Fig4}h). In the $\alpha$-phase, all the observed bands on the (001) surface are also gapped (Fig.~\ref{fig:Fig2}g, h and Fig.~\ref{fig:Fig4}a). A gap of  $\sim$85 meV can be further seen from the bands tracked by the EDC stacks in Fig.~\ref{fig:Fig4}a. On the (100)' surface, the quasi-1D Dirac-like surface state still persists (Fig.~\ref{fig:Fig2}e, f). From a high-resolution measurement taken with 30 eV photons at 10 K (Fig.~\ref{fig:Fig4}c), we observe a gap of $\sim$35 meV clearly in the EDC stacks, indicating a gapped surface state within a bulk band gap. This is consistent with our theoretical prediction and calculation of a higher-order TI scenario of $\alpha$-\bi~\cite{Yoon2020}, illustrated in Fig.~\ref{fig:Fig4}g, where both the bulk bands and surfaces bands are required to be gapped, hosting gapless hinge states inside the surface gap. A comparison of our calculated bulk and surface states with our measured bands show strong agreement (Fig. S11). We summarize our experimental findings and classification of the two phases in Table~\ref{tab:table1}.

\begin{table*}
\caption{\label{tab:table1}%
Summary of the key features on $\alpha$-\bi~and $\beta$-\bi~revealed by our ARPES study}
\begin{ruledtabular}
\begin{tabular}{cccccc}
Structure&Temperature&Bulk bands&(001) surface&(100) surface&Topology \\
\hline
$\alpha$-\bi&$<$295 K&8N&$\sim$85 meV ($\bar{M}$)&$\sim$35 meV ($\bar{\Gamma}$)&Higher-order TI\\

$\beta$-\bi&$>$305 K&4N&$\sim$100 meV ($\bar{M}$)&Gapless&Weak TI\\

\end{tabular}
\end{ruledtabular}
\end{table*}

Finally, we comment on the inconsistencies in the previous two ARPES studies on \bi. In the first report~\cite{Autes2016}, since no quenching was performed, the phase measured at low temperatures was most likely the $\alpha$-phase instead of the claimed $\beta$-phase. The conclusion that this phase is a strong TI was based on the observation of a gapless Dirac crossing on the (001) surface. Here we show that this Dirac band is rather a bulk band, where high resolution measurements reveal a gap (Fig.~\ref{fig:Fig4}a). In the second report with nanoARPES~\cite{Noguchi2019}, the $\beta$-phase was claimed to be stabilized at low temperature by quenching directly from high temperature. The authors ascribed point-like FS to $\alpha$-\bi~and quasi-1D FS to $\beta$-\bi~based on a superposition of signals from the (001) and (100) surfaces. Based on this assumption, they concluded that the $\beta$-phase is a weak TI and the $\alpha$-phase is a trivial insulator. Benefiting from the relatively large cleavable (100) and (001) surfaces free of domains, here we unambiguously show that both phases have similar quasi-1D FS on the (100) side surface, and that the distinction between the two phases is the subtle band doubling in the $\alpha$-phase, which was not resolved in previous studies. Reminiscent of the bulk Peierls distortion in the Su-Schrieffer-Heeger model~\cite{Su1979}, this yields a gap opening in the Dirac surface state on the (100)' surface, which is the prerequisite for the appearance of mid-gap hinge states in $\alpha$-\bi~as a higher-order TI~\cite{Yoon2020}.

\section{Discussions}
Topological properties of a variety of quantum materials have been verified by a combination of ARPES and first-principles calculations during the last decades~\cite{Chen2009, Hsieh2009, Tanaka2012, Xu2012a, Liu2014, Liu2014a, Xu2015, Lv2015, Lv2019, Sobota2020}. For the quasi-1D material \bi, not only does the uncertainty of the theoretical prediction for the topological classification of both $\alpha$-\bi~and $\beta$-\bi~call for experimental exploration, but also the change of the electronic structures between $\beta$-\bi~and $\alpha$-\bi~is subtle such that careful ARPES measurements with considerable energy resolution is necessary. Unambiguously, the $\beta$-phase \bi~is shown to be the first weak TI by our ARPES measurements on large natural cleavable planes of the top and side surfaces, which resolves the controversy of the previous reports~\cite{Autes2016, Noguchi2019}. For $\alpha$-\bi, two distinctions are clearly observed as compared with $\beta$-\bi: band doubling and gap opening in the side surface state. Together with the identification of the $\beta$-phase as a weak TI, the existence of the Dirac-like surface state with a small gap indicates that the $\alpha$-phase is consistent with the identification as a rare higher-order TI~\cite{Yoon2020} instead of the trivial insulator based on symmetry indicators alone~\cite{Zhang2019a, Vergniory2019, Tang2019}. Since the $\beta$-phase has been shown as a weak TI here, each (001) monolayer is a 2D Z$_2$ TI~\cite{Yoon2020}. As the $\alpha$-phase has been shown to have two monolayers per unit cell here, i.e., a 2D trivial insulator, the 3D bulk must be Z$_2$ trivial given the weak interlayer couplings \cite{Fu2007, Liu2016}; in other words, the surface states are gapped. Moreover, for an $\alpha$-phase thin film with a large odd number of monolayers, it can be viewed as a 2D Z$_2$ TI with a helical edge state. Apparently, this edge state can only appear at hinges and form a loop, demonstrating that the α-phase is a higher-order TI \cite{Yoon2020}. However, we do note that further direct evidence of the hinge states is desirable. 

Fundamentally, our result implies that even without strong correlation and magnetism, there are still many topological materials that are beyond the scope of the already efficacious topological quantum chemistry or symmetry indicators~\cite{Zhang2019a, Vergniory2019, Tang2019}. Significantly, our work reveals the strong tunability of the quasi-1D TI \bi~as a novel material platform where the abundance of topological phases could be explored. In the future, it would be exciting to investigate in this platform, for example, the strain-induced topological phases and their transitions, the spatial locations and Luttinger parameters of helical hinge states, the possible intrinsic and extrinsic topological superconductivity, as well as applications of our identified room-temperature topological phase transition as a topological switch that controls the dimensions of the boundary conduction channels.


\noindent {\bf Acknowledgement} This work is mainly supported by National Science Foundation (NSF) through the DMREF program. This research used resources of the Stanford Synchrotron Radiation Lightsource and the Advanced Light Source, both U.S. DOE Office of Science User Facilities under contract Nos. AC02-76SF00515 and DE-AC02-05CH112319, respectively. We acknowledge the Texas Advanced Computing Center (TACC) for providing resources that have contributed to the research results reported in this work. The work at Rice is supported by NSF under Grant No. DMR-1921847, the Gordon and Betty Moore Foundation's EPiQS Initiative through grant no. GBMF9470, the Robert A. Welch Foundation under Grant No. C-2024, and the Sloan Foundation FG-2019-12224. The work at UCB is supported by NSF Grant No. DMR-1921798. The work at UTD is supported by NSF under Grant Nos. DMR-1921581 and DMR-1945351, AFOSR under Grant No. FA9550-19-1-0037, and Army Research Office (ARO) under Grant No. W911NF-18-1-0416. The work at OSU is supported by NSF under Grant No. DMR-1922076.

\bibliography{Bi4I4_reference}

\begin{thebibliography}{40}%
\makeatletter
\providecommand \@ifxundefined [1]{%
 \@ifx{#1\undefined}
}%
\providecommand \@ifnum [1]{%
 \ifnum #1\expandafter \@firstoftwo
 \else \expandafter \@secondoftwo
 \fi
}%
\providecommand \@ifx [1]{%
 \ifx #1\expandafter \@firstoftwo
 \else \expandafter \@secondoftwo
 \fi
}%
\providecommand \natexlab [1]{#1}%
\providecommand \enquote  [1]{``#1''}%
\providecommand \bibnamefont  [1]{#1}%
\providecommand \bibfnamefont [1]{#1}%
\providecommand \citenamefont [1]{#1}%
\providecommand \href@noop [0]{\@secondoftwo}%
\providecommand \href [0]{\begingroup \@sanitize@url \@href}%
\providecommand \@href[1]{\@@startlink{#1}\@@href}%
\providecommand \@@href[1]{\endgroup#1\@@endlink}%
\providecommand \@sanitize@url [0]{\catcode `\\12\catcode `\$12\catcode
  `\&12\catcode `\#12\catcode `\^12\catcode `\_12\catcode `\%12\relax}%
\providecommand \@@startlink[1]{}%
\providecommand \@@endlink[0]{}%
\providecommand \url  [0]{\begingroup\@sanitize@url \@url }%
\providecommand \@url [1]{\endgroup\@href {#1}{\urlprefix }}%
\providecommand \urlprefix  [0]{URL }%
\providecommand \Eprint [0]{\href }%
\providecommand \doibase [0]{https://doi.org/}%
\providecommand \selectlanguage [0]{\@gobble}%
\providecommand \bibinfo  [0]{\@secondoftwo}%
\providecommand \bibfield  [0]{\@secondoftwo}%
\providecommand \translation [1]{[#1]}%
\providecommand \BibitemOpen [0]{}%
\providecommand \bibitemStop [0]{}%
\providecommand \bibitemNoStop [0]{.\EOS\space}%
\providecommand \EOS [0]{\spacefactor3000\relax}%
\providecommand \BibitemShut  [1]{\csname bibitem#1\endcsname}%
\let\auto@bib@innerbib\@empty
\bibitem [{\citenamefont {Kane}\ and\ \citenamefont {Mele}(2005)}]{Kane2005a}%
  \BibitemOpen
  \bibfield  {author} {\bibinfo {author} {\bibfnamefont {C.~L.}\ \bibnamefont
  {Kane}}\ and\ \bibinfo {author} {\bibfnamefont {E.~J.}\ \bibnamefont
  {Mele}},\ }\bibfield  {title} {\bibinfo {title} {{Z$_2$ Topological Order and
  the Quantum Spin Hall Effect}},\ }\href
  {https://doi.org/10.1103/PhysRevLett.95.146802} {\bibfield  {journal}
  {\bibinfo  {journal} {Physical Review Letters}\ }\textbf {\bibinfo {volume}
  {95}},\ \bibinfo {pages} {146802} (\bibinfo {year} {2005})}\BibitemShut
  {NoStop}%
\bibitem [{\citenamefont {Konig}\ \emph {et~al.}(2007)\citenamefont {Konig},
  \citenamefont {Wiedmann}, \citenamefont {Brune}, \citenamefont {Roth},
  \citenamefont {Buhmann}, \citenamefont {Molenkamp}, \citenamefont {Qi},\ and\
  \citenamefont {Zhang}}]{Konig2007}%
  \BibitemOpen
  \bibfield  {author} {\bibinfo {author} {\bibfnamefont {M.}~\bibnamefont
  {Konig}}, \bibinfo {author} {\bibfnamefont {S.}~\bibnamefont {Wiedmann}},
  \bibinfo {author} {\bibfnamefont {C.}~\bibnamefont {Brune}}, \bibinfo
  {author} {\bibfnamefont {A.}~\bibnamefont {Roth}}, \bibinfo {author}
  {\bibfnamefont {H.}~\bibnamefont {Buhmann}}, \bibinfo {author} {\bibfnamefont
  {L.~W.}\ \bibnamefont {Molenkamp}}, \bibinfo {author} {\bibfnamefont {X.-L.}\
  \bibnamefont {Qi}},\ and\ \bibinfo {author} {\bibfnamefont {S.-C.}\
  \bibnamefont {Zhang}},\ }\bibfield  {title} {\bibinfo {title} {{Quantum Spin
  Hall Insulator State in HgTe Quantum Wells}},\ }\href
  {https://doi.org/10.1126/science.1148047} {\bibfield  {journal} {\bibinfo
  {journal} {Science}\ }\textbf {\bibinfo {volume} {318}},\ \bibinfo {pages}
  {766} (\bibinfo {year} {2007})}\BibitemShut {NoStop}%
\bibitem [{\citenamefont {Fu}\ \emph {et~al.}(2007)\citenamefont {Fu},
  \citenamefont {Kane},\ and\ \citenamefont {Mele}}]{Fu2007}%
  \BibitemOpen
  \bibfield  {author} {\bibinfo {author} {\bibfnamefont {L.}~\bibnamefont
  {Fu}}, \bibinfo {author} {\bibfnamefont {C.~L.}\ \bibnamefont {Kane}},\ and\
  \bibinfo {author} {\bibfnamefont {E.~J.}\ \bibnamefont {Mele}},\ }\bibfield
  {title} {\bibinfo {title} {{Topological Insulators in Three Dimensions}},\
  }\href {https://doi.org/10.1103/PhysRevLett.98.106803} {\bibfield  {journal}
  {\bibinfo  {journal} {Physical Review Letters}\ }\textbf {\bibinfo {volume}
  {98}},\ \bibinfo {pages} {106803} (\bibinfo {year} {2007})}\BibitemShut
  {NoStop}%
\bibitem [{\citenamefont {Moore}\ and\ \citenamefont
  {Balents}(2007)}]{Moore2007}%
  \BibitemOpen
  \bibfield  {author} {\bibinfo {author} {\bibfnamefont {J.~E.}\ \bibnamefont
  {Moore}}\ and\ \bibinfo {author} {\bibfnamefont {L.}~\bibnamefont
  {Balents}},\ }\bibfield  {title} {\bibinfo {title} {{Topological invariants
  of time-reversal-invariant band structures}},\ }\href
  {https://doi.org/10.1103/PhysRevB.75.121306} {\bibfield  {journal} {\bibinfo
  {journal} {Physical Review B}\ }\textbf {\bibinfo {volume} {75}},\ \bibinfo
  {pages} {121306} (\bibinfo {year} {2007})}\BibitemShut {NoStop}%
\bibitem [{\citenamefont {Hasan}\ and\ \citenamefont {Kane}(2010)}]{Hasan2010}%
  \BibitemOpen
  \bibfield  {author} {\bibinfo {author} {\bibfnamefont {M.~Z.}\ \bibnamefont
  {Hasan}}\ and\ \bibinfo {author} {\bibfnamefont {C.~L.}\ \bibnamefont
  {Kane}},\ }\bibfield  {title} {\bibinfo {title} {{Colloquium : Topological
  insulators}},\ }\href {https://doi.org/10.1103/RevModPhys.82.3045} {\bibfield
   {journal} {\bibinfo  {journal} {Reviews of Modern Physics}\ }\textbf
  {\bibinfo {volume} {82}},\ \bibinfo {pages} {3045} (\bibinfo {year}
  {2010})}\BibitemShut {NoStop}%
\bibitem [{\citenamefont {Qi}\ and\ \citenamefont {Zhang}(2011)}]{Qi2011}%
  \BibitemOpen
  \bibfield  {author} {\bibinfo {author} {\bibfnamefont {X.-L.}\ \bibnamefont
  {Qi}}\ and\ \bibinfo {author} {\bibfnamefont {S.-C.}\ \bibnamefont {Zhang}},\
  }\bibfield  {title} {\bibinfo {title} {{Topological insulators and
  superconductors}},\ }\href {https://doi.org/10.1103/RevModPhys.83.1057}
  {\bibfield  {journal} {\bibinfo  {journal} {Reviews of Modern Physics}\
  }\textbf {\bibinfo {volume} {83}},\ \bibinfo {pages} {1057} (\bibinfo {year}
  {2011})}\BibitemShut {NoStop}%
\bibitem [{\citenamefont {Liu}\ \emph {et~al.}(2016)\citenamefont {Liu},
  \citenamefont {Zhou}, \citenamefont {Yao},\ and\ \citenamefont
  {Zhang}}]{Liu2016}%
  \BibitemOpen
  \bibfield  {author} {\bibinfo {author} {\bibfnamefont {C.-C.}\ \bibnamefont
  {Liu}}, \bibinfo {author} {\bibfnamefont {J.-J.}\ \bibnamefont {Zhou}},
  \bibinfo {author} {\bibfnamefont {Y.}~\bibnamefont {Yao}},\ and\ \bibinfo
  {author} {\bibfnamefont {F.}~\bibnamefont {Zhang}},\ }\bibfield  {title}
  {\bibinfo {title} {{Weak Topological Insulators and Composite Weyl
  Semimetals: $\beta$-Bi$_4$X$_4$ (X=Br, I)}},\ }\href
  {https://doi.org/10.1103/PhysRevLett.116.066801} {\bibfield  {journal}
  {\bibinfo  {journal} {Physical Review Letters}\ }\textbf {\bibinfo {volume}
  {116}},\ \bibinfo {pages} {066801} (\bibinfo {year} {2016})}\BibitemShut
  {NoStop}%
\bibitem [{\citenamefont {Xia}\ \emph {et~al.}(2009)\citenamefont {Xia},
  \citenamefont {Qian}, \citenamefont {Hsieh}, \citenamefont {Wray},
  \citenamefont {Pal}, \citenamefont {Lin}, \citenamefont {Bansil},
  \citenamefont {Grauer}, \citenamefont {Hor}, \citenamefont {Cava},\ and\
  \citenamefont {Hasan}}]{Xia2009}%
  \BibitemOpen
  \bibfield  {author} {\bibinfo {author} {\bibfnamefont {Y.}~\bibnamefont
  {Xia}}, \bibinfo {author} {\bibfnamefont {D.}~\bibnamefont {Qian}}, \bibinfo
  {author} {\bibfnamefont {D.}~\bibnamefont {Hsieh}}, \bibinfo {author}
  {\bibfnamefont {L.}~\bibnamefont {Wray}}, \bibinfo {author} {\bibfnamefont
  {A.}~\bibnamefont {Pal}}, \bibinfo {author} {\bibfnamefont {H.}~\bibnamefont
  {Lin}}, \bibinfo {author} {\bibfnamefont {A.}~\bibnamefont {Bansil}},
  \bibinfo {author} {\bibfnamefont {D.}~\bibnamefont {Grauer}}, \bibinfo
  {author} {\bibfnamefont {Y.~S.}\ \bibnamefont {Hor}}, \bibinfo {author}
  {\bibfnamefont {R.~J.}\ \bibnamefont {Cava}},\ and\ \bibinfo {author}
  {\bibfnamefont {M.~Z.}\ \bibnamefont {Hasan}},\ }\bibfield  {title} {\bibinfo
  {title} {{Observation of a large-gap topological-insulator class with a
  single Dirac cone on the surface}},\ }\href
  {https://doi.org/10.1038/nphys1274} {\bibfield  {journal} {\bibinfo
  {journal} {Nature Physics}\ }\textbf {\bibinfo {volume} {5}},\ \bibinfo
  {pages} {398} (\bibinfo {year} {2009})}\BibitemShut {NoStop}%
\bibitem [{\citenamefont {Zhang}\ \emph {et~al.}(2009)\citenamefont {Zhang},
  \citenamefont {Liu}, \citenamefont {Qi}, \citenamefont {Dai}, \citenamefont
  {Fang},\ and\ \citenamefont {Zhang}}]{Zhang2009}%
  \BibitemOpen
  \bibfield  {author} {\bibinfo {author} {\bibfnamefont {H.}~\bibnamefont
  {Zhang}}, \bibinfo {author} {\bibfnamefont {C.-X.}\ \bibnamefont {Liu}},
  \bibinfo {author} {\bibfnamefont {X.-L.}\ \bibnamefont {Qi}}, \bibinfo
  {author} {\bibfnamefont {X.}~\bibnamefont {Dai}}, \bibinfo {author}
  {\bibfnamefont {Z.}~\bibnamefont {Fang}},\ and\ \bibinfo {author}
  {\bibfnamefont {S.-C.}\ \bibnamefont {Zhang}},\ }\bibfield  {title} {\bibinfo
  {title} {{Topological insulators in Bi$_2$Se$_3$, Bi$_2$Te$_3$ and
  Sb$_2$Te$_3$ with a single Dirac cone on the surface}},\ }\href
  {https://doi.org/10.1038/nphys1270} {\bibfield  {journal} {\bibinfo
  {journal} {Nature Physics}\ }\textbf {\bibinfo {volume} {5}},\ \bibinfo
  {pages} {438} (\bibinfo {year} {2009})}\BibitemShut {NoStop}%
\bibitem [{\citenamefont {Chen}\ \emph {et~al.}(2009)\citenamefont {Chen},
  \citenamefont {Analytis}, \citenamefont {Chu}, \citenamefont {Liu},
  \citenamefont {Mo}, \citenamefont {Qi}, \citenamefont {Zhang}, \citenamefont
  {Lu}, \citenamefont {Dai}, \citenamefont {Fang}, \citenamefont {Zhang},
  \citenamefont {Fisher}, \citenamefont {Hussain},\ and\ \citenamefont
  {Shen}}]{Chen2009}%
  \BibitemOpen
  \bibfield  {author} {\bibinfo {author} {\bibfnamefont {Y.~L.}\ \bibnamefont
  {Chen}}, \bibinfo {author} {\bibfnamefont {J.~G.}\ \bibnamefont {Analytis}},
  \bibinfo {author} {\bibfnamefont {J.-H.}\ \bibnamefont {Chu}}, \bibinfo
  {author} {\bibfnamefont {Z.~K.}\ \bibnamefont {Liu}}, \bibinfo {author}
  {\bibfnamefont {S.-K.}\ \bibnamefont {Mo}}, \bibinfo {author} {\bibfnamefont
  {X.~L.}\ \bibnamefont {Qi}}, \bibinfo {author} {\bibfnamefont {H.~J.}\
  \bibnamefont {Zhang}}, \bibinfo {author} {\bibfnamefont {D.~H.}\ \bibnamefont
  {Lu}}, \bibinfo {author} {\bibfnamefont {X.}~\bibnamefont {Dai}}, \bibinfo
  {author} {\bibfnamefont {Z.}~\bibnamefont {Fang}}, \bibinfo {author}
  {\bibfnamefont {S.~C.}\ \bibnamefont {Zhang}}, \bibinfo {author}
  {\bibfnamefont {I.~R.}\ \bibnamefont {Fisher}}, \bibinfo {author}
  {\bibfnamefont {Z.}~\bibnamefont {Hussain}},\ and\ \bibinfo {author}
  {\bibfnamefont {Z.-X.}\ \bibnamefont {Shen}},\ }\bibfield  {title} {\bibinfo
  {title} {{Experimental Realization of a Three-Dimensional Topological
  Insulator, Bi$_2$Te$_3$}},\ }\href {https://doi.org/10.1126/science.1173034}
  {\bibfield  {journal} {\bibinfo  {journal} {Science}\ }\textbf {\bibinfo
  {volume} {325}},\ \bibinfo {pages} {178} (\bibinfo {year}
  {2009})}\BibitemShut {NoStop}%
\bibitem [{\citenamefont {Hsieh}\ \emph {et~al.}(2009)\citenamefont {Hsieh},
  \citenamefont {Xia}, \citenamefont {Qian}, \citenamefont {Wray},
  \citenamefont {Dil}, \citenamefont {Meier}, \citenamefont {Osterwalder},
  \citenamefont {Patthey}, \citenamefont {Checkelsky}, \citenamefont {Ong},
  \citenamefont {Fedorov}, \citenamefont {Lin}, \citenamefont {Bansil},
  \citenamefont {Grauer}, \citenamefont {Hor}, \citenamefont {Cava},\ and\
  \citenamefont {Hasan}}]{Hsieh2009}%
  \BibitemOpen
  \bibfield  {author} {\bibinfo {author} {\bibfnamefont {D.}~\bibnamefont
  {Hsieh}}, \bibinfo {author} {\bibfnamefont {Y.}~\bibnamefont {Xia}}, \bibinfo
  {author} {\bibfnamefont {D.}~\bibnamefont {Qian}}, \bibinfo {author}
  {\bibfnamefont {L.}~\bibnamefont {Wray}}, \bibinfo {author} {\bibfnamefont
  {J.~H.}\ \bibnamefont {Dil}}, \bibinfo {author} {\bibfnamefont
  {F.}~\bibnamefont {Meier}}, \bibinfo {author} {\bibfnamefont
  {J.}~\bibnamefont {Osterwalder}}, \bibinfo {author} {\bibfnamefont
  {L.}~\bibnamefont {Patthey}}, \bibinfo {author} {\bibfnamefont {J.~G.}\
  \bibnamefont {Checkelsky}}, \bibinfo {author} {\bibfnamefont {N.~P.}\
  \bibnamefont {Ong}}, \bibinfo {author} {\bibfnamefont {A.~V.}\ \bibnamefont
  {Fedorov}}, \bibinfo {author} {\bibfnamefont {H.}~\bibnamefont {Lin}},
  \bibinfo {author} {\bibfnamefont {A.}~\bibnamefont {Bansil}}, \bibinfo
  {author} {\bibfnamefont {D.}~\bibnamefont {Grauer}}, \bibinfo {author}
  {\bibfnamefont {Y.~S.}\ \bibnamefont {Hor}}, \bibinfo {author} {\bibfnamefont
  {R.~J.}\ \bibnamefont {Cava}},\ and\ \bibinfo {author} {\bibfnamefont
  {M.~Z.}\ \bibnamefont {Hasan}},\ }\bibfield  {title} {\bibinfo {title} {{A
  tunable topological insulator in the spin helical Dirac transport regime}},\
  }\href {https://doi.org/10.1038/nature08234} {\bibfield  {journal} {\bibinfo
  {journal} {Nature}\ }\textbf {\bibinfo {volume} {460}},\ \bibinfo {pages}
  {1101} (\bibinfo {year} {2009})}\BibitemShut {NoStop}%
\bibitem [{\citenamefont {Yan}\ \emph {et~al.}(2012)\citenamefont {Yan},
  \citenamefont {M{\"{u}}chler},\ and\ \citenamefont {Felser}}]{Yan2012}%
  \BibitemOpen
  \bibfield  {author} {\bibinfo {author} {\bibfnamefont {B.}~\bibnamefont
  {Yan}}, \bibinfo {author} {\bibfnamefont {L.}~\bibnamefont {M{\"{u}}chler}},\
  and\ \bibinfo {author} {\bibfnamefont {C.}~\bibnamefont {Felser}},\
  }\bibfield  {title} {\bibinfo {title} {{Prediction of Weak Topological
  Insulators in Layered Semiconductors}},\ }\href
  {https://doi.org/10.1103/PhysRevLett.109.116406} {\bibfield  {journal}
  {\bibinfo  {journal} {Physical Review Letters}\ }\textbf {\bibinfo {volume}
  {109}},\ \bibinfo {pages} {116406} (\bibinfo {year} {2012})}\BibitemShut
  {NoStop}%
\bibitem [{\citenamefont {Rasche}\ \emph {et~al.}(2013)\citenamefont {Rasche},
  \citenamefont {Isaeva}, \citenamefont {Ruck}, \citenamefont {Borisenko},
  \citenamefont {Zabolotnyy}, \citenamefont {B{\"{u}}chner}, \citenamefont
  {Koepernik}, \citenamefont {Ortix}, \citenamefont {Richter},\ and\
  \citenamefont {van~den Brink}}]{Rasche2013}%
  \BibitemOpen
  \bibfield  {author} {\bibinfo {author} {\bibfnamefont {B.}~\bibnamefont
  {Rasche}}, \bibinfo {author} {\bibfnamefont {A.}~\bibnamefont {Isaeva}},
  \bibinfo {author} {\bibfnamefont {M.}~\bibnamefont {Ruck}}, \bibinfo {author}
  {\bibfnamefont {S.}~\bibnamefont {Borisenko}}, \bibinfo {author}
  {\bibfnamefont {V.}~\bibnamefont {Zabolotnyy}}, \bibinfo {author}
  {\bibfnamefont {B.}~\bibnamefont {B{\"{u}}chner}}, \bibinfo {author}
  {\bibfnamefont {K.}~\bibnamefont {Koepernik}}, \bibinfo {author}
  {\bibfnamefont {C.}~\bibnamefont {Ortix}}, \bibinfo {author} {\bibfnamefont
  {M.}~\bibnamefont {Richter}},\ and\ \bibinfo {author} {\bibfnamefont
  {J.}~\bibnamefont {van~den Brink}},\ }\bibfield  {title} {\bibinfo {title}
  {{Stacked topological insulator built from bismuth-based graphene sheet
  analogues}},\ }\href {https://doi.org/10.1038/nmat3570} {\bibfield  {journal}
  {\bibinfo  {journal} {Nature Materials}\ }\textbf {\bibinfo {volume} {12}},\
  \bibinfo {pages} {422} (\bibinfo {year} {2013})}\BibitemShut {NoStop}%
\bibitem [{\citenamefont {Tang}\ \emph {et~al.}(2014)\citenamefont {Tang},
  \citenamefont {Yan}, \citenamefont {Cao}, \citenamefont {Wu}, \citenamefont
  {Felser},\ and\ \citenamefont {Duan}}]{Tang2014}%
  \BibitemOpen
  \bibfield  {author} {\bibinfo {author} {\bibfnamefont {P.}~\bibnamefont
  {Tang}}, \bibinfo {author} {\bibfnamefont {B.}~\bibnamefont {Yan}}, \bibinfo
  {author} {\bibfnamefont {W.}~\bibnamefont {Cao}}, \bibinfo {author}
  {\bibfnamefont {S.-C.}\ \bibnamefont {Wu}}, \bibinfo {author} {\bibfnamefont
  {C.}~\bibnamefont {Felser}},\ and\ \bibinfo {author} {\bibfnamefont
  {W.}~\bibnamefont {Duan}},\ }\bibfield  {title} {\bibinfo {title} {{Weak
  topological insulators induced by the interlayer coupling: A first-principles
  study of stacked Bi$_2$TeI}},\ }\href
  {https://doi.org/10.1103/PhysRevB.89.041409} {\bibfield  {journal} {\bibinfo
  {journal} {Physical Review B}\ }\textbf {\bibinfo {volume} {89}},\ \bibinfo
  {pages} {041409} (\bibinfo {year} {2014})}\BibitemShut {NoStop}%
\bibitem [{\citenamefont {Yang}\ \emph {et~al.}(2014)\citenamefont {Yang},
  \citenamefont {Liu}, \citenamefont {Fu}, \citenamefont {Duan},\ and\
  \citenamefont {Liu}}]{Yang2014}%
  \BibitemOpen
  \bibfield  {author} {\bibinfo {author} {\bibfnamefont {G.}~\bibnamefont
  {Yang}}, \bibinfo {author} {\bibfnamefont {J.}~\bibnamefont {Liu}}, \bibinfo
  {author} {\bibfnamefont {L.}~\bibnamefont {Fu}}, \bibinfo {author}
  {\bibfnamefont {W.}~\bibnamefont {Duan}},\ and\ \bibinfo {author}
  {\bibfnamefont {C.}~\bibnamefont {Liu}},\ }\bibfield  {title} {\bibinfo
  {title} {{Weak topological insulators in PbTe/SnTe superlattices}},\ }\href
  {https://doi.org/10.1103/PhysRevB.89.085312} {\bibfield  {journal} {\bibinfo
  {journal} {Physical Review B}\ }\textbf {\bibinfo {volume} {89}},\ \bibinfo
  {pages} {085312} (\bibinfo {year} {2014})}\BibitemShut {NoStop}%
\bibitem [{\citenamefont {Li}\ \emph {et~al.}(2015)\citenamefont {Li},
  \citenamefont {Zhang}, \citenamefont {Niu},\ and\ \citenamefont
  {Feng}}]{Li2015}%
  \BibitemOpen
  \bibfield  {author} {\bibinfo {author} {\bibfnamefont {X.}~\bibnamefont
  {Li}}, \bibinfo {author} {\bibfnamefont {F.}~\bibnamefont {Zhang}}, \bibinfo
  {author} {\bibfnamefont {Q.}~\bibnamefont {Niu}},\ and\ \bibinfo {author}
  {\bibfnamefont {J.}~\bibnamefont {Feng}},\ }\bibfield  {title} {\bibinfo
  {title} {{Superlattice valley engineering for designer topological
  insulators}},\ }\href {https://doi.org/10.1038/srep06397} {\bibfield
  {journal} {\bibinfo  {journal} {Scientific Reports}\ }\textbf {\bibinfo
  {volume} {4}},\ \bibinfo {pages} {6397} (\bibinfo {year} {2015})}\BibitemShut
  {NoStop}%
\bibitem [{\citenamefont {Zhang}\ \emph {et~al.}(2013)\citenamefont {Zhang},
  \citenamefont {Kane},\ and\ \citenamefont {Mele}}]{Zhang2013}%
  \BibitemOpen
  \bibfield  {author} {\bibinfo {author} {\bibfnamefont {F.}~\bibnamefont
  {Zhang}}, \bibinfo {author} {\bibfnamefont {C.~L.}\ \bibnamefont {Kane}},\
  and\ \bibinfo {author} {\bibfnamefont {E.~J.}\ \bibnamefont {Mele}},\
  }\bibfield  {title} {\bibinfo {title} {{Surface State Magnetization and
  Chiral Edge States on Topological Insulators}},\ }\href
  {https://doi.org/10.1103/PhysRevLett.110.046404} {\bibfield  {journal}
  {\bibinfo  {journal} {Physical Review Letters}\ }\textbf {\bibinfo {volume}
  {110}},\ \bibinfo {pages} {046404} (\bibinfo {year} {2013})}\BibitemShut
  {NoStop}%
\bibitem [{\citenamefont {Po}\ \emph {et~al.}(2017)\citenamefont {Po},
  \citenamefont {Vishwanath},\ and\ \citenamefont {Watanabe}}]{Po2017}%
  \BibitemOpen
  \bibfield  {author} {\bibinfo {author} {\bibfnamefont {H.~C.}\ \bibnamefont
  {Po}}, \bibinfo {author} {\bibfnamefont {A.}~\bibnamefont {Vishwanath}},\
  and\ \bibinfo {author} {\bibfnamefont {H.}~\bibnamefont {Watanabe}},\
  }\bibfield  {title} {\bibinfo {title} {{Symmetry-based indicators of band
  topology in the 230 space groups}},\ }\href
  {https://doi.org/10.1038/s41467-017-00133-2} {\bibfield  {journal} {\bibinfo
  {journal} {Nature Communications}\ }\textbf {\bibinfo {volume} {8}},\
  \bibinfo {pages} {50} (\bibinfo {year} {2017})}\BibitemShut {NoStop}%
\bibitem [{\citenamefont {Schindler}\ \emph {et~al.}(2018)\citenamefont
  {Schindler}, \citenamefont {Cook}, \citenamefont {Vergniory}, \citenamefont
  {Wang}, \citenamefont {Parkin}, \citenamefont {Bernevig},\ and\ \citenamefont
  {Neupert}}]{Schindler2018}%
  \BibitemOpen
  \bibfield  {author} {\bibinfo {author} {\bibfnamefont {F.}~\bibnamefont
  {Schindler}}, \bibinfo {author} {\bibfnamefont {A.~M.}\ \bibnamefont {Cook}},
  \bibinfo {author} {\bibfnamefont {M.~G.}\ \bibnamefont {Vergniory}}, \bibinfo
  {author} {\bibfnamefont {Z.}~\bibnamefont {Wang}}, \bibinfo {author}
  {\bibfnamefont {S.~S.~P.}\ \bibnamefont {Parkin}}, \bibinfo {author}
  {\bibfnamefont {B.~A.}\ \bibnamefont {Bernevig}},\ and\ \bibinfo {author}
  {\bibfnamefont {T.}~\bibnamefont {Neupert}},\ }\bibfield  {title} {\bibinfo
  {title} {{Higher-order topological insulators}},\ }\href
  {https://doi.org/10.1126/sciadv.aat0346} {\bibfield  {journal} {\bibinfo
  {journal} {Science Advances}\ }\textbf {\bibinfo {volume} {4}},\ \bibinfo
  {pages} {eaat0346} (\bibinfo {year} {2018})}\BibitemShut {NoStop}%
\bibitem [{\citenamefont {Yoon}\ \emph {et~al.}(2020)\citenamefont {Yoon},
  \citenamefont {Liu}, \citenamefont {Min},\ and\ \citenamefont
  {Zhang}}]{Yoon2020}%
  \BibitemOpen
  \bibfield  {author} {\bibinfo {author} {\bibfnamefont {C.}~\bibnamefont
  {Yoon}}, \bibinfo {author} {\bibfnamefont {C.-C.}\ \bibnamefont {Liu}},
  \bibinfo {author} {\bibfnamefont {H.}~\bibnamefont {Min}},\ and\ \bibinfo
  {author} {\bibfnamefont {F.}~\bibnamefont {Zhang}},\ }\bibfield  {title}
  {\bibinfo {title} {{Quasi-One-Dimensional Higher-Order Topological
  Insulators}},\ }\href {http://arxiv.org/abs/2005.14710} {\bibfield  {journal}
  {\bibinfo  {journal} {arXiv}\ }\textbf {\bibinfo {volume} {2005}},\ \bibinfo
  {pages} {14710} (\bibinfo {year} {2020})}\BibitemShut {NoStop}%
\bibitem [{\citenamefont {Weng}\ \emph {et~al.}(2014)\citenamefont {Weng},
  \citenamefont {Dai},\ and\ \citenamefont {Fang}}]{Weng2014}%
  \BibitemOpen
  \bibfield  {author} {\bibinfo {author} {\bibfnamefont {H.}~\bibnamefont
  {Weng}}, \bibinfo {author} {\bibfnamefont {X.}~\bibnamefont {Dai}},\ and\
  \bibinfo {author} {\bibfnamefont {Z.}~\bibnamefont {Fang}},\ }\bibfield
  {title} {\bibinfo {title} {{Transition-Metal Pentatelluride ZrTe$_5$ and
  HfTe$_5$: A paradigm for large-gap quantum spin hall insulators}},\ }\href
  {https://doi.org/10.1103/PhysRevX.4.011002} {\bibfield  {journal} {\bibinfo
  {journal} {Physical Review X}\ }\textbf {\bibinfo {volume} {4}},\ \bibinfo
  {pages} {011002} (\bibinfo {year} {2014})}\BibitemShut {NoStop}%
\bibitem [{\citenamefont {Mutch}\ \emph {et~al.}(2019)\citenamefont {Mutch},
  \citenamefont {Chen}, \citenamefont {Went}, \citenamefont {Qian},
  \citenamefont {Wilson}, \citenamefont {Andreev}, \citenamefont {Chen},\ and\
  \citenamefont {Chu}}]{Mutch2019}%
  \BibitemOpen
  \bibfield  {author} {\bibinfo {author} {\bibfnamefont {J.}~\bibnamefont
  {Mutch}}, \bibinfo {author} {\bibfnamefont {W.-C.}\ \bibnamefont {Chen}},
  \bibinfo {author} {\bibfnamefont {P.}~\bibnamefont {Went}}, \bibinfo {author}
  {\bibfnamefont {T.}~\bibnamefont {Qian}}, \bibinfo {author} {\bibfnamefont
  {I.~Z.}\ \bibnamefont {Wilson}}, \bibinfo {author} {\bibfnamefont
  {A.}~\bibnamefont {Andreev}}, \bibinfo {author} {\bibfnamefont {C.-C.}\
  \bibnamefont {Chen}},\ and\ \bibinfo {author} {\bibfnamefont {J.-H.}\
  \bibnamefont {Chu}},\ }\bibfield  {title} {\bibinfo {title} {{Evidence for a
  strain-tuned topological phase transition in ZrTe$_5$}},\ }\href
  {https://doi.org/10.1126/sciadv.aav9771} {\bibfield  {journal} {\bibinfo
  {journal} {Science Advances}\ }\textbf {\bibinfo {volume} {5}},\ \bibinfo
  {pages} {eaav9771} (\bibinfo {year} {2019})}\BibitemShut {NoStop}%
\bibitem [{\citenamefont {Zhang}\ \emph {et~al.}(2021)\citenamefont {Zhang},
  \citenamefont {Noguchi}, \citenamefont {Kuroda}, \citenamefont {Lin},
  \citenamefont {Kawaguchi}, \citenamefont {Yaji}, \citenamefont {Harasawa},
  \citenamefont {Lippmaa}, \citenamefont {Nie}, \citenamefont {Weng},
  \citenamefont {Kandyba}, \citenamefont {Giampietri}, \citenamefont {Barinov},
  \citenamefont {Li}, \citenamefont {Gu}, \citenamefont {Shin},\ and\
  \citenamefont {Kondo}}]{Zhang2021}%
  \BibitemOpen
  \bibfield  {author} {\bibinfo {author} {\bibfnamefont {P.}~\bibnamefont
  {Zhang}}, \bibinfo {author} {\bibfnamefont {R.}~\bibnamefont {Noguchi}},
  \bibinfo {author} {\bibfnamefont {K.}~\bibnamefont {Kuroda}}, \bibinfo
  {author} {\bibfnamefont {C.}~\bibnamefont {Lin}}, \bibinfo {author}
  {\bibfnamefont {K.}~\bibnamefont {Kawaguchi}}, \bibinfo {author}
  {\bibfnamefont {K.}~\bibnamefont {Yaji}}, \bibinfo {author} {\bibfnamefont
  {A.}~\bibnamefont {Harasawa}}, \bibinfo {author} {\bibfnamefont
  {M.}~\bibnamefont {Lippmaa}}, \bibinfo {author} {\bibfnamefont
  {S.}~\bibnamefont {Nie}}, \bibinfo {author} {\bibfnamefont {H.}~\bibnamefont
  {Weng}}, \bibinfo {author} {\bibfnamefont {V.}~\bibnamefont {Kandyba}},
  \bibinfo {author} {\bibfnamefont {A.}~\bibnamefont {Giampietri}}, \bibinfo
  {author} {\bibfnamefont {A.}~\bibnamefont {Barinov}}, \bibinfo {author}
  {\bibfnamefont {Q.}~\bibnamefont {Li}}, \bibinfo {author} {\bibfnamefont
  {G.~D.}\ \bibnamefont {Gu}}, \bibinfo {author} {\bibfnamefont
  {S.}~\bibnamefont {Shin}},\ and\ \bibinfo {author} {\bibfnamefont
  {T.}~\bibnamefont {Kondo}},\ }\bibfield  {title} {\bibinfo {title}
  {{Observation and control of the weak topological insulator state in
  ZrTe$_5$}},\ }\href {https://doi.org/10.1038/s41467-020-20564-8} {\bibfield
  {journal} {\bibinfo  {journal} {Nature Communications}\ }\textbf {\bibinfo
  {volume} {12}},\ \bibinfo {pages} {406} (\bibinfo {year} {2021})}\BibitemShut
  {NoStop}%
\bibitem [{\citenamefont {von Schnering}\ \emph {et~al.}(1978)\citenamefont
  {von Schnering}, \citenamefont {von Benda},\ and\ \citenamefont
  {Kalveram}}]{VonSchnering1978}%
  \BibitemOpen
  \bibfield  {author} {\bibinfo {author} {\bibfnamefont {H.~G.}\ \bibnamefont
  {von Schnering}}, \bibinfo {author} {\bibfnamefont {H.}~\bibnamefont {von
  Benda}},\ and\ \bibinfo {author} {\bibfnamefont {C.}~\bibnamefont
  {Kalveram}},\ }\bibfield  {title} {\bibinfo {title} {{Wismutmonojodid BiJ,
  eine Verbindung mit Bi(O) und Bi(II)}},\ }\href
  {https://doi.org/10.1002/zaac.19784380104} {\bibfield  {journal} {\bibinfo
  {journal} {Zeitschrift f{\"{u}}r anorganische und allgemeine Chemie}\
  }\textbf {\bibinfo {volume} {438}},\ \bibinfo {pages} {37} (\bibinfo {year}
  {1978})}\BibitemShut {NoStop}%
\bibitem [{\citenamefont {Aut{\`{e}}s}\ \emph {et~al.}(2016)\citenamefont
  {Aut{\`{e}}s}, \citenamefont {Isaeva}, \citenamefont {Moreschini},
  \citenamefont {Johannsen}, \citenamefont {Pisoni}, \citenamefont {Mori},
  \citenamefont {Zhang}, \citenamefont {Filatova}, \citenamefont {Kuznetsov},
  \citenamefont {Forr{\'{o}}}, \citenamefont {{Van den Broek}}, \citenamefont
  {Kim}, \citenamefont {Kim}, \citenamefont {Lanzara}, \citenamefont
  {Denlinger}, \citenamefont {Rotenberg}, \citenamefont {Bostwick},
  \citenamefont {Grioni},\ and\ \citenamefont {Yazyev}}]{Autes2016}%
  \BibitemOpen
  \bibfield  {author} {\bibinfo {author} {\bibfnamefont {G.}~\bibnamefont
  {Aut{\`{e}}s}}, \bibinfo {author} {\bibfnamefont {A.}~\bibnamefont {Isaeva}},
  \bibinfo {author} {\bibfnamefont {L.}~\bibnamefont {Moreschini}}, \bibinfo
  {author} {\bibfnamefont {J.~C.}\ \bibnamefont {Johannsen}}, \bibinfo {author}
  {\bibfnamefont {A.}~\bibnamefont {Pisoni}}, \bibinfo {author} {\bibfnamefont
  {R.}~\bibnamefont {Mori}}, \bibinfo {author} {\bibfnamefont {W.}~\bibnamefont
  {Zhang}}, \bibinfo {author} {\bibfnamefont {T.~G.}\ \bibnamefont {Filatova}},
  \bibinfo {author} {\bibfnamefont {A.~N.}\ \bibnamefont {Kuznetsov}}, \bibinfo
  {author} {\bibfnamefont {L.}~\bibnamefont {Forr{\'{o}}}}, \bibinfo {author}
  {\bibfnamefont {W.}~\bibnamefont {{Van den Broek}}}, \bibinfo {author}
  {\bibfnamefont {Y.}~\bibnamefont {Kim}}, \bibinfo {author} {\bibfnamefont
  {K.~S.}\ \bibnamefont {Kim}}, \bibinfo {author} {\bibfnamefont
  {A.}~\bibnamefont {Lanzara}}, \bibinfo {author} {\bibfnamefont {J.~D.}\
  \bibnamefont {Denlinger}}, \bibinfo {author} {\bibfnamefont {E.}~\bibnamefont
  {Rotenberg}}, \bibinfo {author} {\bibfnamefont {A.}~\bibnamefont {Bostwick}},
  \bibinfo {author} {\bibfnamefont {M.}~\bibnamefont {Grioni}},\ and\ \bibinfo
  {author} {\bibfnamefont {O.~V.}\ \bibnamefont {Yazyev}},\ }\bibfield  {title}
  {\bibinfo {title} {{A novel quasi-one-dimensional topological insulator in
  bismuth iodide $\beta$-Bi$_4$I$_4$}},\ }\href
  {https://doi.org/10.1038/nmat4488} {\bibfield  {journal} {\bibinfo  {journal}
  {Nature Materials}\ }\textbf {\bibinfo {volume} {15}},\ \bibinfo {pages}
  {154} (\bibinfo {year} {2016})}\BibitemShut {NoStop}%
\bibitem [{\citenamefont {Noguchi}\ \emph {et~al.}(2019)\citenamefont
  {Noguchi}, \citenamefont {Takahashi}, \citenamefont {Kuroda}, \citenamefont
  {Ochi}, \citenamefont {Shirasawa}, \citenamefont {Sakano}, \citenamefont
  {Bareille}, \citenamefont {Nakayama}, \citenamefont {Watson}, \citenamefont
  {Yaji}, \citenamefont {Harasawa}, \citenamefont {Iwasawa}, \citenamefont
  {Dudin}, \citenamefont {Kim}, \citenamefont {Hoesch}, \citenamefont
  {Kandyba}, \citenamefont {Giampietri}, \citenamefont {Barinov}, \citenamefont
  {Shin}, \citenamefont {Arita}, \citenamefont {Sasagawa},\ and\ \citenamefont
  {Kondo}}]{Noguchi2019}%
  \BibitemOpen
  \bibfield  {author} {\bibinfo {author} {\bibfnamefont {R.}~\bibnamefont
  {Noguchi}}, \bibinfo {author} {\bibfnamefont {T.}~\bibnamefont {Takahashi}},
  \bibinfo {author} {\bibfnamefont {K.}~\bibnamefont {Kuroda}}, \bibinfo
  {author} {\bibfnamefont {M.}~\bibnamefont {Ochi}}, \bibinfo {author}
  {\bibfnamefont {T.}~\bibnamefont {Shirasawa}}, \bibinfo {author}
  {\bibfnamefont {M.}~\bibnamefont {Sakano}}, \bibinfo {author} {\bibfnamefont
  {C.}~\bibnamefont {Bareille}}, \bibinfo {author} {\bibfnamefont
  {M.}~\bibnamefont {Nakayama}}, \bibinfo {author} {\bibfnamefont {M.~D.}\
  \bibnamefont {Watson}}, \bibinfo {author} {\bibfnamefont {K.}~\bibnamefont
  {Yaji}}, \bibinfo {author} {\bibfnamefont {A.}~\bibnamefont {Harasawa}},
  \bibinfo {author} {\bibfnamefont {H.}~\bibnamefont {Iwasawa}}, \bibinfo
  {author} {\bibfnamefont {P.}~\bibnamefont {Dudin}}, \bibinfo {author}
  {\bibfnamefont {T.~K.}\ \bibnamefont {Kim}}, \bibinfo {author} {\bibfnamefont
  {M.}~\bibnamefont {Hoesch}}, \bibinfo {author} {\bibfnamefont
  {V.}~\bibnamefont {Kandyba}}, \bibinfo {author} {\bibfnamefont
  {A.}~\bibnamefont {Giampietri}}, \bibinfo {author} {\bibfnamefont
  {A.}~\bibnamefont {Barinov}}, \bibinfo {author} {\bibfnamefont
  {S.}~\bibnamefont {Shin}}, \bibinfo {author} {\bibfnamefont {R.}~\bibnamefont
  {Arita}}, \bibinfo {author} {\bibfnamefont {T.}~\bibnamefont {Sasagawa}},\
  and\ \bibinfo {author} {\bibfnamefont {T.}~\bibnamefont {Kondo}},\ }\bibfield
   {title} {\bibinfo {title} {{A weak topological insulator state in
  quasi-one-dimensional bismuth iodide}},\ }\href
  {https://doi.org/10.1038/s41586-019-0927-7} {\bibfield  {journal} {\bibinfo
  {journal} {Nature}\ }\textbf {\bibinfo {volume} {566}},\ \bibinfo {pages}
  {518} (\bibinfo {year} {2019})}\BibitemShut {NoStop}%
\bibitem [{\citenamefont {Zhang}\ \emph {et~al.}(2019)\citenamefont {Zhang},
  \citenamefont {Jiang}, \citenamefont {Song}, \citenamefont {Huang},
  \citenamefont {He}, \citenamefont {Fang}, \citenamefont {Weng},\ and\
  \citenamefont {Fang}}]{Zhang2019a}%
  \BibitemOpen
  \bibfield  {author} {\bibinfo {author} {\bibfnamefont {T.}~\bibnamefont
  {Zhang}}, \bibinfo {author} {\bibfnamefont {Y.}~\bibnamefont {Jiang}},
  \bibinfo {author} {\bibfnamefont {Z.}~\bibnamefont {Song}}, \bibinfo {author}
  {\bibfnamefont {H.}~\bibnamefont {Huang}}, \bibinfo {author} {\bibfnamefont
  {Y.}~\bibnamefont {He}}, \bibinfo {author} {\bibfnamefont {Z.}~\bibnamefont
  {Fang}}, \bibinfo {author} {\bibfnamefont {H.}~\bibnamefont {Weng}},\ and\
  \bibinfo {author} {\bibfnamefont {C.}~\bibnamefont {Fang}},\ }\bibfield
  {title} {\bibinfo {title} {{Catalogue of topological electronic materials}},\
  }\href {https://doi.org/10.1038/s41586-019-0944-6} {\bibfield  {journal}
  {\bibinfo  {journal} {Nature}\ }\textbf {\bibinfo {volume} {566}},\ \bibinfo
  {pages} {475} (\bibinfo {year} {2019})}\BibitemShut {NoStop}%
\bibitem [{\citenamefont {Vergniory}\ \emph {et~al.}(2019)\citenamefont
  {Vergniory}, \citenamefont {Elcoro}, \citenamefont {Felser}, \citenamefont
  {Regnault}, \citenamefont {Bernevig},\ and\ \citenamefont
  {Wang}}]{Vergniory2019}%
  \BibitemOpen
  \bibfield  {author} {\bibinfo {author} {\bibfnamefont {M.~G.}\ \bibnamefont
  {Vergniory}}, \bibinfo {author} {\bibfnamefont {L.}~\bibnamefont {Elcoro}},
  \bibinfo {author} {\bibfnamefont {C.}~\bibnamefont {Felser}}, \bibinfo
  {author} {\bibfnamefont {N.}~\bibnamefont {Regnault}}, \bibinfo {author}
  {\bibfnamefont {B.~A.}\ \bibnamefont {Bernevig}},\ and\ \bibinfo {author}
  {\bibfnamefont {Z.}~\bibnamefont {Wang}},\ }\bibfield  {title} {\bibinfo
  {title} {{A complete catalogue of high-quality topological materials}},\
  }\href {https://doi.org/10.1038/s41586-019-0954-4} {\bibfield  {journal}
  {\bibinfo  {journal} {Nature}\ }\textbf {\bibinfo {volume} {566}},\ \bibinfo
  {pages} {480} (\bibinfo {year} {2019})}\BibitemShut {NoStop}%
\bibitem [{\citenamefont {Tang}\ \emph
  {et~al.}(2019{\natexlab{a}})\citenamefont {Tang}, \citenamefont {Po},
  \citenamefont {Vishwanath},\ and\ \citenamefont {Wan}}]{Tang2019}%
  \BibitemOpen
  \bibfield  {author} {\bibinfo {author} {\bibfnamefont {F.}~\bibnamefont
  {Tang}}, \bibinfo {author} {\bibfnamefont {H.~C.}\ \bibnamefont {Po}},
  \bibinfo {author} {\bibfnamefont {A.}~\bibnamefont {Vishwanath}},\ and\
  \bibinfo {author} {\bibfnamefont {X.}~\bibnamefont {Wan}},\ }\bibfield
  {title} {\bibinfo {title} {{Comprehensive search for topological materials
  using symmetry indicators}},\ }\href
  {https://doi.org/10.1038/s41586-019-0937-5} {\bibfield  {journal} {\bibinfo
  {journal} {Nature}\ }\textbf {\bibinfo {volume} {566}},\ \bibinfo {pages}
  {486} (\bibinfo {year} {2019}{\natexlab{a}})}\BibitemShut {NoStop}%
\bibitem [{\citenamefont {Halperin}(1987)}]{Halperin1987a}%
  \BibitemOpen
  \bibfield  {author} {\bibinfo {author} {\bibfnamefont {B.~I.}\ \bibnamefont
  {Halperin}},\ }\bibfield  {title} {\bibinfo {title} {{Possible states for a
  three-dimensional electron gas in a strong magnetic field}},\ }\href
  {https://doi.org/10.7567/JJAPS.26S3.1913} {\bibfield  {journal} {\bibinfo
  {journal} {Japanese Journal of Applied Physics}\ }\textbf {\bibinfo {volume}
  {26}},\ \bibinfo {pages} {1913} (\bibinfo {year} {1987})}\BibitemShut
  {NoStop}%
\bibitem [{\citenamefont {Tang}\ \emph
  {et~al.}(2019{\natexlab{b}})\citenamefont {Tang}, \citenamefont {Ren},
  \citenamefont {Wang}, \citenamefont {Zhong}, \citenamefont {Schneeloch},
  \citenamefont {Yang}, \citenamefont {Yang}, \citenamefont {Lee},
  \citenamefont {Gu}, \citenamefont {Qiao},\ and\ \citenamefont
  {Zhang}}]{Tang2019a}%
  \BibitemOpen
  \bibfield  {author} {\bibinfo {author} {\bibfnamefont {F.}~\bibnamefont
  {Tang}}, \bibinfo {author} {\bibfnamefont {Y.}~\bibnamefont {Ren}}, \bibinfo
  {author} {\bibfnamefont {P.}~\bibnamefont {Wang}}, \bibinfo {author}
  {\bibfnamefont {R.}~\bibnamefont {Zhong}}, \bibinfo {author} {\bibfnamefont
  {J.}~\bibnamefont {Schneeloch}}, \bibinfo {author} {\bibfnamefont {S.~A.}\
  \bibnamefont {Yang}}, \bibinfo {author} {\bibfnamefont {K.}~\bibnamefont
  {Yang}}, \bibinfo {author} {\bibfnamefont {P.~A.}\ \bibnamefont {Lee}},
  \bibinfo {author} {\bibfnamefont {G.}~\bibnamefont {Gu}}, \bibinfo {author}
  {\bibfnamefont {Z.}~\bibnamefont {Qiao}},\ and\ \bibinfo {author}
  {\bibfnamefont {L.}~\bibnamefont {Zhang}},\ }\bibfield  {title} {\bibinfo
  {title} {{Three-dimensional quantum Hall effect and metal–insulator
  transition in ZrTe$_5$}},\ }\href {https://doi.org/10.1038/s41586-019-1180-9}
  {\bibfield  {journal} {\bibinfo  {journal} {Nature}\ }\textbf {\bibinfo
  {volume} {569}},\ \bibinfo {pages} {537} (\bibinfo {year}
  {2019}{\natexlab{b}})}\BibitemShut {NoStop}%
\bibitem [{\citenamefont {Su}\ \emph {et~al.}(1979)\citenamefont {Su},
  \citenamefont {Schrieffer},\ and\ \citenamefont {Heeger}}]{Su1979}%
  \BibitemOpen
  \bibfield  {author} {\bibinfo {author} {\bibfnamefont {W.~P.}\ \bibnamefont
  {Su}}, \bibinfo {author} {\bibfnamefont {J.~R.}\ \bibnamefont {Schrieffer}},\
  and\ \bibinfo {author} {\bibfnamefont {A.~J.}\ \bibnamefont {Heeger}},\
  }\bibfield  {title} {\bibinfo {title} {{Solitons in Polyacetylene}},\ }\href
  {https://doi.org/10.1103/PhysRevLett.42.1698} {\bibfield  {journal} {\bibinfo
   {journal} {Physical Review Letters}\ }\textbf {\bibinfo {volume} {42}},\
  \bibinfo {pages} {1698} (\bibinfo {year} {1979})}\BibitemShut {NoStop}%
\bibitem [{\citenamefont {Tanaka}\ \emph {et~al.}(2012)\citenamefont {Tanaka},
  \citenamefont {Ren}, \citenamefont {Sato}, \citenamefont {Nakayama},
  \citenamefont {Souma}, \citenamefont {Takahashi}, \citenamefont {Segawa},\
  and\ \citenamefont {Ando}}]{Tanaka2012}%
  \BibitemOpen
  \bibfield  {author} {\bibinfo {author} {\bibfnamefont {Y.}~\bibnamefont
  {Tanaka}}, \bibinfo {author} {\bibfnamefont {Z.}~\bibnamefont {Ren}},
  \bibinfo {author} {\bibfnamefont {T.}~\bibnamefont {Sato}}, \bibinfo {author}
  {\bibfnamefont {K.}~\bibnamefont {Nakayama}}, \bibinfo {author}
  {\bibfnamefont {S.}~\bibnamefont {Souma}}, \bibinfo {author} {\bibfnamefont
  {T.}~\bibnamefont {Takahashi}}, \bibinfo {author} {\bibfnamefont
  {K.}~\bibnamefont {Segawa}},\ and\ \bibinfo {author} {\bibfnamefont
  {Y.}~\bibnamefont {Ando}},\ }\bibfield  {title} {\bibinfo {title}
  {{Experimental realization of a topological crystalline insulator in SnTe}},\
  }\href {https://doi.org/10.1038/nphys2442} {\bibfield  {journal} {\bibinfo
  {journal} {Nature Physics}\ }\textbf {\bibinfo {volume} {8}},\ \bibinfo
  {pages} {800} (\bibinfo {year} {2012})}\BibitemShut {NoStop}%
\bibitem [{\citenamefont {Xu}\ \emph {et~al.}(2012)\citenamefont {Xu},
  \citenamefont {Liu}, \citenamefont {Alidoust}, \citenamefont {Neupane},
  \citenamefont {Qian}, \citenamefont {Belopolski}, \citenamefont {Denlinger},
  \citenamefont {Wang}, \citenamefont {Lin}, \citenamefont {Wray},
  \citenamefont {Landolt}, \citenamefont {Slomski}, \citenamefont {Dil},
  \citenamefont {Marcinkova}, \citenamefont {Morosan}, \citenamefont {Gibson},
  \citenamefont {Sankar}, \citenamefont {Chou}, \citenamefont {Cava},
  \citenamefont {Bansil},\ and\ \citenamefont {Hasan}}]{Xu2012a}%
  \BibitemOpen
  \bibfield  {author} {\bibinfo {author} {\bibfnamefont {S.-Y.}\ \bibnamefont
  {Xu}}, \bibinfo {author} {\bibfnamefont {C.}~\bibnamefont {Liu}}, \bibinfo
  {author} {\bibfnamefont {N.}~\bibnamefont {Alidoust}}, \bibinfo {author}
  {\bibfnamefont {M.}~\bibnamefont {Neupane}}, \bibinfo {author} {\bibfnamefont
  {D.}~\bibnamefont {Qian}}, \bibinfo {author} {\bibfnamefont {I.}~\bibnamefont
  {Belopolski}}, \bibinfo {author} {\bibfnamefont {J.}~\bibnamefont
  {Denlinger}}, \bibinfo {author} {\bibfnamefont {Y.}~\bibnamefont {Wang}},
  \bibinfo {author} {\bibfnamefont {H.}~\bibnamefont {Lin}}, \bibinfo {author}
  {\bibfnamefont {L.}~\bibnamefont {Wray}}, \bibinfo {author} {\bibfnamefont
  {G.}~\bibnamefont {Landolt}}, \bibinfo {author} {\bibfnamefont
  {B.}~\bibnamefont {Slomski}}, \bibinfo {author} {\bibfnamefont
  {J.}~\bibnamefont {Dil}}, \bibinfo {author} {\bibfnamefont {A.}~\bibnamefont
  {Marcinkova}}, \bibinfo {author} {\bibfnamefont {E.}~\bibnamefont {Morosan}},
  \bibinfo {author} {\bibfnamefont {Q.}~\bibnamefont {Gibson}}, \bibinfo
  {author} {\bibfnamefont {R.}~\bibnamefont {Sankar}}, \bibinfo {author}
  {\bibfnamefont {F.}~\bibnamefont {Chou}}, \bibinfo {author} {\bibfnamefont
  {R.}~\bibnamefont {Cava}}, \bibinfo {author} {\bibfnamefont {A.}~\bibnamefont
  {Bansil}},\ and\ \bibinfo {author} {\bibfnamefont {M.}~\bibnamefont
  {Hasan}},\ }\bibfield  {title} {\bibinfo {title} {{Observation of a
  topological crystalline insulator phase and topological phase transition in
  Pb$_{1-x}$Sn$_x$Te}},\ }\href {https://doi.org/10.1038/ncomms2191} {\bibfield
   {journal} {\bibinfo  {journal} {Nature Communications}\ }\textbf {\bibinfo
  {volume} {3}},\ \bibinfo {pages} {1192} (\bibinfo {year} {2012})}\BibitemShut
  {NoStop}%
\bibitem [{\citenamefont {Liu}\ \emph {et~al.}(2014{\natexlab{a}})\citenamefont
  {Liu}, \citenamefont {Jiang}, \citenamefont {Zhou}, \citenamefont {Wang},
  \citenamefont {Zhang}, \citenamefont {Weng}, \citenamefont {Prabhakaran},
  \citenamefont {Mo}, \citenamefont {Peng}, \citenamefont {Dudin},
  \citenamefont {Kim}, \citenamefont {Hoesch}, \citenamefont {Fang},
  \citenamefont {Dai}, \citenamefont {Shen}, \citenamefont {Feng},
  \citenamefont {Hussain},\ and\ \citenamefont {Chen}}]{Liu2014}%
  \BibitemOpen
  \bibfield  {author} {\bibinfo {author} {\bibfnamefont {Z.~K.}\ \bibnamefont
  {Liu}}, \bibinfo {author} {\bibfnamefont {J.}~\bibnamefont {Jiang}}, \bibinfo
  {author} {\bibfnamefont {B.}~\bibnamefont {Zhou}}, \bibinfo {author}
  {\bibfnamefont {Z.~J.}\ \bibnamefont {Wang}}, \bibinfo {author}
  {\bibfnamefont {Y.}~\bibnamefont {Zhang}}, \bibinfo {author} {\bibfnamefont
  {H.~M.}\ \bibnamefont {Weng}}, \bibinfo {author} {\bibfnamefont
  {D.}~\bibnamefont {Prabhakaran}}, \bibinfo {author} {\bibfnamefont {S.-K.}\
  \bibnamefont {Mo}}, \bibinfo {author} {\bibfnamefont {H.}~\bibnamefont
  {Peng}}, \bibinfo {author} {\bibfnamefont {P.}~\bibnamefont {Dudin}},
  \bibinfo {author} {\bibfnamefont {T.}~\bibnamefont {Kim}}, \bibinfo {author}
  {\bibfnamefont {M.}~\bibnamefont {Hoesch}}, \bibinfo {author} {\bibfnamefont
  {Z.}~\bibnamefont {Fang}}, \bibinfo {author} {\bibfnamefont {X.}~\bibnamefont
  {Dai}}, \bibinfo {author} {\bibfnamefont {Z.~X.}\ \bibnamefont {Shen}},
  \bibinfo {author} {\bibfnamefont {D.~L.}\ \bibnamefont {Feng}}, \bibinfo
  {author} {\bibfnamefont {Z.}~\bibnamefont {Hussain}},\ and\ \bibinfo {author}
  {\bibfnamefont {Y.~L.}\ \bibnamefont {Chen}},\ }\bibfield  {title} {\bibinfo
  {title} {{A stable three-dimensional topological Dirac semimetal
  Cd$_3$As$_2$}},\ }\href {https://doi.org/10.1038/nmat3990} {\bibfield
  {journal} {\bibinfo  {journal} {Nature Materials}\ }\textbf {\bibinfo
  {volume} {13}},\ \bibinfo {pages} {677} (\bibinfo {year}
  {2014}{\natexlab{a}})}\BibitemShut {NoStop}%
\bibitem [{\citenamefont {Liu}\ \emph {et~al.}(2014{\natexlab{b}})\citenamefont
  {Liu}, \citenamefont {Zhou}, \citenamefont {Zhang}, \citenamefont {Wang},
  \citenamefont {Weng}, \citenamefont {Prabhakaran}, \citenamefont {Mo},
  \citenamefont {Shen}, \citenamefont {Fang}, \citenamefont {Dai},
  \citenamefont {Hussain},\ and\ \citenamefont {Chen}}]{Liu2014a}%
  \BibitemOpen
  \bibfield  {author} {\bibinfo {author} {\bibfnamefont {Z.~K.}\ \bibnamefont
  {Liu}}, \bibinfo {author} {\bibfnamefont {B.}~\bibnamefont {Zhou}}, \bibinfo
  {author} {\bibfnamefont {Y.}~\bibnamefont {Zhang}}, \bibinfo {author}
  {\bibfnamefont {Z.~J.}\ \bibnamefont {Wang}}, \bibinfo {author}
  {\bibfnamefont {H.~M.}\ \bibnamefont {Weng}}, \bibinfo {author}
  {\bibfnamefont {D.}~\bibnamefont {Prabhakaran}}, \bibinfo {author}
  {\bibfnamefont {S.-K.}\ \bibnamefont {Mo}}, \bibinfo {author} {\bibfnamefont
  {Z.~X.}\ \bibnamefont {Shen}}, \bibinfo {author} {\bibfnamefont
  {Z.}~\bibnamefont {Fang}}, \bibinfo {author} {\bibfnamefont {X.}~\bibnamefont
  {Dai}}, \bibinfo {author} {\bibfnamefont {Z.}~\bibnamefont {Hussain}},\ and\
  \bibinfo {author} {\bibfnamefont {Y.~L.}\ \bibnamefont {Chen}},\ }\bibfield
  {title} {\bibinfo {title} {{Discovery of a Three-Dimensional Topological
  Dirac Semimetal, Na$_3$Bi}},\ }\href
  {https://doi.org/10.1126/science.1245085} {\bibfield  {journal} {\bibinfo
  {journal} {Science}\ }\textbf {\bibinfo {volume} {343}},\ \bibinfo {pages}
  {864} (\bibinfo {year} {2014}{\natexlab{b}})}\BibitemShut {NoStop}%
\bibitem [{\citenamefont {Xu}\ \emph {et~al.}(2015)\citenamefont {Xu},
  \citenamefont {Belopolski}, \citenamefont {Alidoust}, \citenamefont
  {Neupane}, \citenamefont {Bian}, \citenamefont {Zhang}, \citenamefont
  {Sankar}, \citenamefont {Chang}, \citenamefont {Yuan}, \citenamefont {Lee},
  \citenamefont {Huang}, \citenamefont {Zheng}, \citenamefont {Ma},
  \citenamefont {Sanchez}, \citenamefont {Wang}, \citenamefont {Bansil},
  \citenamefont {Chou}, \citenamefont {Shibayev}, \citenamefont {Lin},
  \citenamefont {Jia},\ and\ \citenamefont {Hasan}}]{Xu2015}%
  \BibitemOpen
  \bibfield  {author} {\bibinfo {author} {\bibfnamefont {S.-Y.}\ \bibnamefont
  {Xu}}, \bibinfo {author} {\bibfnamefont {I.}~\bibnamefont {Belopolski}},
  \bibinfo {author} {\bibfnamefont {N.}~\bibnamefont {Alidoust}}, \bibinfo
  {author} {\bibfnamefont {M.}~\bibnamefont {Neupane}}, \bibinfo {author}
  {\bibfnamefont {G.}~\bibnamefont {Bian}}, \bibinfo {author} {\bibfnamefont
  {C.}~\bibnamefont {Zhang}}, \bibinfo {author} {\bibfnamefont
  {R.}~\bibnamefont {Sankar}}, \bibinfo {author} {\bibfnamefont
  {G.}~\bibnamefont {Chang}}, \bibinfo {author} {\bibfnamefont
  {Z.}~\bibnamefont {Yuan}}, \bibinfo {author} {\bibfnamefont {C.-C.}\
  \bibnamefont {Lee}}, \bibinfo {author} {\bibfnamefont {S.-M.}\ \bibnamefont
  {Huang}}, \bibinfo {author} {\bibfnamefont {H.}~\bibnamefont {Zheng}},
  \bibinfo {author} {\bibfnamefont {J.}~\bibnamefont {Ma}}, \bibinfo {author}
  {\bibfnamefont {D.~S.}\ \bibnamefont {Sanchez}}, \bibinfo {author}
  {\bibfnamefont {B.}~\bibnamefont {Wang}}, \bibinfo {author} {\bibfnamefont
  {A.}~\bibnamefont {Bansil}}, \bibinfo {author} {\bibfnamefont
  {F.}~\bibnamefont {Chou}}, \bibinfo {author} {\bibfnamefont {P.~P.}\
  \bibnamefont {Shibayev}}, \bibinfo {author} {\bibfnamefont {H.}~\bibnamefont
  {Lin}}, \bibinfo {author} {\bibfnamefont {S.}~\bibnamefont {Jia}},\ and\
  \bibinfo {author} {\bibfnamefont {M.~Z.}\ \bibnamefont {Hasan}},\ }\bibfield
  {title} {\bibinfo {title} {{Discovery of a Weyl fermion semimetal and
  topological Fermi arcs}},\ }\href {https://doi.org/10.1126/science.aaa9297}
  {\bibfield  {journal} {\bibinfo  {journal} {Science}\ }\textbf {\bibinfo
  {volume} {349}},\ \bibinfo {pages} {613} (\bibinfo {year}
  {2015})}\BibitemShut {NoStop}%
\bibitem [{\citenamefont {Lv}\ \emph {et~al.}(2015)\citenamefont {Lv},
  \citenamefont {Xu}, \citenamefont {Weng}, \citenamefont {Ma}, \citenamefont
  {Richard}, \citenamefont {Huang}, \citenamefont {Zhao}, \citenamefont {Chen},
  \citenamefont {Matt}, \citenamefont {Bisti}, \citenamefont {Strocov},
  \citenamefont {Mesot}, \citenamefont {Fang}, \citenamefont {Dai},
  \citenamefont {Qian}, \citenamefont {Shi},\ and\ \citenamefont
  {Ding}}]{Lv2015}%
  \BibitemOpen
  \bibfield  {author} {\bibinfo {author} {\bibfnamefont {B.~Q.}\ \bibnamefont
  {Lv}}, \bibinfo {author} {\bibfnamefont {N.}~\bibnamefont {Xu}}, \bibinfo
  {author} {\bibfnamefont {H.~M.}\ \bibnamefont {Weng}}, \bibinfo {author}
  {\bibfnamefont {J.~Z.}\ \bibnamefont {Ma}}, \bibinfo {author} {\bibfnamefont
  {P.}~\bibnamefont {Richard}}, \bibinfo {author} {\bibfnamefont {X.~C.}\
  \bibnamefont {Huang}}, \bibinfo {author} {\bibfnamefont {L.~X.}\ \bibnamefont
  {Zhao}}, \bibinfo {author} {\bibfnamefont {G.~F.}\ \bibnamefont {Chen}},
  \bibinfo {author} {\bibfnamefont {C.~E.}\ \bibnamefont {Matt}}, \bibinfo
  {author} {\bibfnamefont {F.}~\bibnamefont {Bisti}}, \bibinfo {author}
  {\bibfnamefont {V.~N.}\ \bibnamefont {Strocov}}, \bibinfo {author}
  {\bibfnamefont {J.}~\bibnamefont {Mesot}}, \bibinfo {author} {\bibfnamefont
  {Z.}~\bibnamefont {Fang}}, \bibinfo {author} {\bibfnamefont {X.}~\bibnamefont
  {Dai}}, \bibinfo {author} {\bibfnamefont {T.}~\bibnamefont {Qian}}, \bibinfo
  {author} {\bibfnamefont {M.}~\bibnamefont {Shi}},\ and\ \bibinfo {author}
  {\bibfnamefont {H.}~\bibnamefont {Ding}},\ }\bibfield  {title} {\bibinfo
  {title} {{Observation of Weyl nodes in TaAs}},\ }\href
  {https://doi.org/10.1038/nphys3426} {\bibfield  {journal} {\bibinfo
  {journal} {Nature Physics}\ }\textbf {\bibinfo {volume} {11}},\ \bibinfo
  {pages} {724} (\bibinfo {year} {2015})}\BibitemShut {NoStop}%
\bibitem [{\citenamefont {Lv}\ \emph {et~al.}(2019)\citenamefont {Lv},
  \citenamefont {Qian},\ and\ \citenamefont {Ding}}]{Lv2019}%
  \BibitemOpen
  \bibfield  {author} {\bibinfo {author} {\bibfnamefont {B.}~\bibnamefont
  {Lv}}, \bibinfo {author} {\bibfnamefont {T.}~\bibnamefont {Qian}},\ and\
  \bibinfo {author} {\bibfnamefont {H.}~\bibnamefont {Ding}},\ }\bibfield
  {title} {\bibinfo {title} {{Angle-resolved photoemission spectroscopy and its
  application to topological materials}},\ }\href
  {https://doi.org/10.1038/s42254-019-0088-5} {\bibfield  {journal} {\bibinfo
  {journal} {Nature Reviews Physics}\ }\textbf {\bibinfo {volume} {1}},\
  \bibinfo {pages} {609} (\bibinfo {year} {2019})}\BibitemShut {NoStop}%
\bibitem [{\citenamefont {Sobota}\ \emph {et~al.}(2020)\citenamefont {Sobota},
  \citenamefont {He},\ and\ \citenamefont {Shen}}]{Sobota2020}%
  \BibitemOpen
  \bibfield  {author} {\bibinfo {author} {\bibfnamefont {J.~A.}\ \bibnamefont
  {Sobota}}, \bibinfo {author} {\bibfnamefont {Y.}~\bibnamefont {He}},\ and\
  \bibinfo {author} {\bibfnamefont {Z.-X.}\ \bibnamefont {Shen}},\ }\bibfield
  {title} {\bibinfo {title} {{Electronic structure of quantum materials studied
  by angle-resolved photoemission spectroscopy}},\ }\href
  {http://arxiv.org/abs/2008.02378} {\bibfield  {journal} {\bibinfo  {journal}
  {arXiv}\ }\textbf {\bibinfo {volume} {2008}},\ \bibinfo {pages} {02378}
  (\bibinfo {year} {2020})}\BibitemShut {NoStop}%
\end{thebibliography}%

\end{document}